\documentclass[twocolumn,superscriptaddress,aps]{revtex4}
\usepackage{amsxtra,amssymb,bm,amsbsy,amsmath,amsthm,latexsym,dsfont,array,layout,graphicx,physics,mathrsfs,color,soul,subfigure,enumerate,footmisc,scrextend}
\usepackage[colorlinks=true,citecolor=blue,linkcolor=blue,anchorcolor=blue,urlcolor=blue]{hyperref}
\hypersetup{linktoc=all}

\newcommand{\blk}{\color{black}}

\begin{document}

\title{Controlling the dynamics of open quantum systems with periodic driving field}

\author{Fei-Lei Xiong}
\affiliation{Department of Physics and Center for Quantum Information Science,
National Cheng Kung University, Tainan 70101, Taiwan}

\author{Wei-Min Zhang}
\email{wzhang@mail.ncku.edu.tw}
\affiliation{Department of Physics and Center for Quantum Information Science,
National Cheng Kung University, Tainan 70101, Taiwan}

\begin{abstract}
In this paper, we study the exact dynamics of open quantum systems to the case with periodic driving field.  %The decoherence dynamics is studied with both the analytic and the numerical methods. 
It is shown that different from the static adjustment of the system on-site energy that can either generate or destroy the dissipationless 
localized bound states, the periodic driving can either preserve the existed localized bound states or destroy some of them 
but cannot generate new localized bound states. With the picture of energy transfer involved with \blk  the driving field, 
we find the condition for the survival of \blk the localized bound states when the driving amplitude is weak. 
For the strong driving case, the condition breaks down because of the strong energy renormalization to the originally existed localized bound states. 
These properties of decoherence dynamics may help in controlling the quantum state against decoherence for the sake of its 
sensitivity to the fundamental frequency of the driving field. 
\end{abstract}

\maketitle

\section{Introduction}

Quantum decoherence acts as a key obstacle in implementing the novel proposals of quantum information science and 
technology, such as quantum computation~\cite{D00}, quantum metrology~\cite{C81,WBI92,CHP12} and quantum simulation~\cite{BSA12}.  Because it is ubiquitous in all kinds of physical devices, studying, reducing, and controlling decoherence have been a long-standing hot topic in these fields.
%Since the 1980s, much effort has been put into it and rich results have been %obtained~\cite{dA17,Z19,CL83,HPZ92,ZLX12,TZ08,LZ12,XZ20}. 
After decades of study, the understanding of decoherence is now much deeper \cite{dA17,Z19}. 
Physicists have gone far beyond the 
Born-Markov master equation~\cite{V54}, the GKS-Lindblad master equation~\cite{GKS76,L76}, and truncated 
Nakajima-Zwanzig master equation~\cite{N58,Z60,BP02}, which are either only applicable in the weak-coupling 
regime or just an approximation to the practically unsolvable formal equation. 
% Memory effects and their connection with the properties of open quantum systems have been heavily studied~\cite{dA17,LHW18,Z19}. 
Up to now, the exact master equations for quantum Brownian motion~\cite{CL83,HPZ92}, for electronic and photonic  
open quantum systems~\cite{ZLX12,TZ08,LZ12,XZ20} and for hybrid topological superconducting systems~\cite{LYH18,HYZ20}, \blk
as well as those that are mathematically equivalent to them, have been successfully derived. 

In the past ten years, we have systematically studied a large class of open quantum system incorporating  
quantum transport (see the reviews~\cite{YZ17,Z19} \blk and the references therein), and obtained the exact master equations as 
well as their solutions expressed in terms of the non-equilibrium Green's functions~\cite{ZLX12,TZ08,LZ12,Z19,XZ20}. 
It is proved that the existence of dissipationless localized bound states (localized modes) which are induced from 
the structured system-environment interactions ~\cite{ZLX12,Z19} offers us a deep understanding of how the decoherence 
of open quantum systems can be suppressed by these localized bound states. They form a decoherence-free subspace without the additional requirement of symmetry\blk, and therefore may have
the potential of storing quantum information in the protocols of quantum information processing.  

On the other hand, as for the controlling of decoherence, ever since the proposal of spin-echo~\cite{H50,CP54}, many methods, e.g., dynamical coupling~\cite{VL98,VK99,Z00,BL02,WBL02} and introducing spatial periodicity~\cite{PCZ96,EFW05,FTT05,JGB11,LMY11,MC15} have been developed.  %For instance, the decoherence suppression protocols based on pulses such as the dynamical coupling~\cite{VL98,VK99,Z00} and leakage elimination operators~\cite{BL02,WBL02,BLWZ05,ZHB20}. 
%for two-level systems such as electrons and spins, experimentalists tend to apply the dynamical coupling~\cite{VL98,VK99,Z00}. For continuous-variable systems, such as quantum harmonic oscillators, physicists have proposed the scheme based on leakage elimination operators~\cite{BL02,WBL02,BLWZ05,ZHB20}. 
%Recently, physicists find that other than imposing time-dependent pulses to the system, physicists also tried to design the environmental structure, e.g., introducing spatial periodicity, in order to generate static bound states which naturally fulfill the aim of suppressing decoherence~\cite{PCZ96,EFW05,FTT05,JGB11,LMY11,MC15}. 
Other than the above-listed protocols, it has also long been confirmed that temporal periodic driving on the open system can impose significant effects on the system dynamics~\cite{GDJ91,GH98,ZYL09,XWZ12,LLY14,OK19,CZ21}. Some methods of dealing with this problem, e.g.,  spectral filtering theory~\cite{KK01,U07,CLN08,BDU11,GSU13} and those based on the Floquet theory~\cite{CAL15,MWA18,BCW21}, have also been developed. 
%It is shown that dissipationless bound states in the total system may be possibly generated, which may fulfill the aim of suppressing decoherence. 
Different from the bound states generated in the time-independent system, e.g., the impurity system in the solid system, the bound states generated by temporal periodic driving fields are time-dependent and are dubbed as Floquet bound states~\cite{CAL15,MWA18,BCW21}. The conditions for their formation as well as their  applications in the quantum information technologies are also discussed~\cite{CAL15,BCW21}.

% In the past ten years,
%By tuning the on-site energy of the system, the localized bound states can be generated or be destroyed, depending on the specific parameters. 
%In our previous works, we have systematically studied transporting systems incorporating with just particle exchange (See the review~\cite{Z19} and the references therein). Exact master equations as well as the solutions have been derived and expressed in terms of the non-equilibrium Green's functions~\cite{ZLX12,TZ08,LZ12,Z19,XZ20}. It was proved that the existence of localized bound states resides in the spectral density property of the open systems~\cite{ZLX12}. Such localized bound states are dissipationless and possess the potential of storing quantum information~\cite{CZ21}. 
In this paper, we shall extend our systematic works~\cite{ZLX12,TZ08,LZ12,Z19,XZ20} on the open quantum systems 
with the existence of localized bund states %incorporating with only particle exchanges 
to the case that a time-periodic driving field is exerted to the system.  In this work, we shall analyze the influences of the driving field and propose the potential application of the driving based on the numerical results and theoretical analysis. The framework is based on the picture of energy transfer and renormalization effect originating \blk from the driving field, which is different from those based on the Floquet theory~\cite{CAL15,BCW21,OK19,MWA18}.
The paper is organized as follows. In Sec.~\ref{sec2}, we introduce the system we study and the general framework for the later discussions. In Sec.~\ref{sec3}, we numerically study the dynamical properties of the system under weak periodic drive. In Sec.~\ref{sec4}, it is shown that when the driving field becomes strong, the dynamical property would be significantly different. The conclusion and the potential applications are given in Sec.~\ref{sec5}.

\section{The general formalism}
\label{sec2}

We shall focus on a class of fermionic open quantum systems driven by a periodic external field. The total Hamiltonian can be modeled as
\begin{align}\label{eq_H}
H_{\rm tot}(t)=\, & [\epsilon_s +  \epsilon_d(t)] b^\dag b +\sum_{\alpha k} \epsilon_{\alpha k} b_{\alpha k}^\dag 
b_{\alpha k}  \nonumber\\ 
& +\sum_{\alpha k} (V_{\alpha k} b^\dag b_{\alpha k} + {h.c.})\,,
%&+ \epsilon_d(t) b^\dag b \,,
\end{align}
where the first term is the system Hamiltonian, $\epsilon_s$ is the system on-site energy, 
$\epsilon_d(t)$ is an external periodic driving field with fundamental period $T$, and
$b$ ($b^\dag$) is the system annihilation (creation) operator. The second term is the environment Hamiltonian, $\epsilon_{\alpha k}$ is the single-particle energy of the mode-$k$ in the reservoir $\alpha$, 
with $b_{\alpha k}$ and $b_{\alpha k}^\dag$ standing for its annihilation and creation operator, respectively. 
The last term is the system-reservoir coupling with $V_{\alpha k}$ standing for the coupling strength between the system and mode-$k$ of the reservoir $\alpha$.
\iffalse
second line of the \emph{r.h.s.} describes the total Hamiltonian in absence of the driving field and the 
effect of the driving, respectively. described by the Hamiltonian~\cite{TZ08,LZ12,ZLX12}
\begin{align}\label{eq_H}
H=\epsilon_s b^\dag b \!+\!\sum_k \epsilon_k b_k^\dag b_k \!+\!\sum_k (V_{k} b^\dag b_k \!+\!V_{k}^* b_k^\dag b)\,,
\end{align}
and , with the effect summarized as  % time-dependent Hamiltonian $H_d(t)$, which reads
\begin{align}
H_d(t)=\epsilon_d(t) b^\dag b\,.
\end{align}
To control the system dynamics, we assume to impose a driving field on the system. 
That is, a time-dependent Hamiltonian $H_d(t)$, which reads
\begin{align}
H_d(t)=\epsilon_d(t) b^\dag b\,,
\end{align}
\fi 
%where $\epsilon_d(t)$ is a function with fundamental period $T$, is added to the original total Hamiltonian. 
Equation \eqref{eq_H} is a prototype in mesoscopic physics that describes quantum transport of a single quantum 
dot coupled to several leads, and the driving field $\epsilon_d(t)$ can be conveniently implemented by exerting the 
time-dependent gate voltage to the quantum dot~\cite{HJ08,YZ17}. 

The decoherence dynamics of the system is largely determined by the behavior of the spectral Green function $u(t,t_0)\!=\!\{b(t),b^\dag(t_0)\}$~\cite{ZLX12}, where $b(t)$ and $b^\dag(t_0)$ are operators %the system annihilation and creation operator 
in the Heisenberg picture and $\{\cdot,\cdot\}$ stands for their anticommutator. Following the equation of motion approach, ${u}(t,t_0)$ satisfies the integro-differential equation~\cite{TZ08}
\begin{align}\label{eq_u}
\frac{d}{d t}{u}(t,t_0)\!+\!i [\epsilon_s\!+\!\epsilon_d(t)] u(t,t_0)\!+\!\int_{t_0}^t d\tau g(t,\tau) u(\tau,t_0)\!=\!0\,,
\end{align}
associated to \blk the initial condition that $u(t_0,t_0)\!=\!1$. Here, $g(t,\tau) \!=\!\int\frac{d\epsilon}{2\pi}
{J}(\epsilon)e^{-i\epsilon(t-\tau)}$ denotes the system-bath correlation \blk, with $J(\epsilon) \!=\! 2\pi \sum_{\alpha k} |V_{\alpha k}|^2 \delta(\epsilon - \epsilon_{\alpha k})$ standing for the spectral density of the system. How the decoherence dynamics is controlled by the driving field is completely determined by the integro-differential equation (\ref{eq_u}).

The periodic drive $\epsilon_d(t)$ can be separated as two parts, one is of the average strength and the other is time-dependent and of zero-average, i.e., \blk %. To express it with the mathematical formula, % $\epsilon_d(t)$ can be written as
\begin{align}
\epsilon_d(t)=\overline{\epsilon}_d\!+\!\Delta\epsilon_d(t) \,,
\end{align}
with $\overline{\epsilon}_d\!=\!\frac{1}{T}\!\int_t^{t\!+\!T}\!\!d\tau\epsilon_d(\tau)$. Actually, the part with strength $\overline{\epsilon}_d$ can be simply counted as the system on-site energy modulation. Without the part $\Delta\epsilon_d(t)$, the tuning of $\overline{\epsilon}_d$ can make the open system generate localized bound states or make the 
existed localized bound state(s) disappear for structure spectral density~\cite{ZLX12}. 
%In another word, by increasing or decreasing the system on-site energy, some single-particle energy level(s) of the total system may be lifted~\cite{FH10} outside the continuous energy band and one or several bound states can be generated; or vice versa, energy level outside the energy band drops into the band and the corresponding bound state is transformed to the scattering state. To summarize,
In another word, the impact of the static part of the driving on the decoherence may be constructive or destructive. Which decoherence effect occurs depends on the parameters setting. In the following sections, we shall demonstrate that the part evolving as 
$\Delta\epsilon_d(t)$ plays an intrinsically different role in modulating the system dynamics in comparison with that given by $\overline{\epsilon}_d$.

We denote that for vanishing $\Delta\epsilon_d(t)$, the spectral Green function is ${u}_0(t,t_0)$. 
\iffalse
Following Eq.~\eqref{eq_u}, it satisfies the equation
\begin{align}
\dot{{u}}_0(t,t_0)\!+\!i (\epsilon_s\!+\!\overline{\epsilon}_d ){u}_0(t,t_0)\!+\!\int_{t_0}^t d\tau g(t,\tau) {u}_0(\tau,t_0)\!=\!0\,,
\end{align}
with the initial condition ${u}_0(t_0,t_0)\!=\!1$.
\fi 
This case \blk has been fully studied in the literature and $u_0(t,t_0)$ is shown to possess the analytic form~\cite{ZLX12}
\iffalse
\begin{align}
u_0(t,t_0)=
\end{align}
\fi 
\begin{align}\label{eq_u_conti}
u_{0}(t,t_0)%=\langle \{c(t), b^\dag(0)\}\rangle
\!=\!\int \frac{d \epsilon}{2\pi} D (\epsilon) e^{-i \epsilon (t-t_0)}\,,
\end{align}
with
\begin{align}\label{eq_dpec}
D(\epsilon)\!=\!2\pi\!\sum_i Z_{i} \delta(\epsilon-\epsilon_{l_i}) % \nonumber\\&
\!+\!\frac{J(\epsilon)}{[\epsilon\!-\!(\epsilon_s+\overline{\epsilon}_d)\!-\!\Delta(\epsilon)]^2 \!+\!J^2(\epsilon)/4}\,. 
\end{align}
%standing for the renormalized spectrum of the system after the coupling effect to the environment has been taken into account. 
In Eq.~\eqref{eq_dpec},  the first term is the localized bound state contribution arisen from the coupling between the system 
and the environment if it exists. Namely, if the condition 
\begin{align}
\epsilon_{l_i}\!-\!(\epsilon_s+\overline{\epsilon}_d)\!-\!\Delta(\epsilon_{l_i}) =0~, ~~ J(\epsilon_{l_i})=0  \label{clbs}
\end{align}
 is satisfied, where  $\Delta(\epsilon)$ is the energy shift from the self-energy correction to the system, $\Sigma(\epsilon)\!=\! \int \frac{d\epsilon'}{2\pi}\frac{J(\epsilon')}{\epsilon-\epsilon'}\!=\!\Delta(\epsilon)-\frac{i}{2} J(\epsilon)$, and
$Z_{i} = [1-\Sigma'(\epsilon_{l_i}]^{-1}$ is the amplitude of the $i$th localized bound state. The existence of localized bound state can be manipulated  
by the modulation of the system on-site energy through the static part $\overline{\epsilon}_d$ of the driving field such that Eq.~(\ref{clbs}) can be obeyed.  
%$J(\epsilon)$ and $\Delta(\epsilon)$ are the spectral density and self-energy of the open system, respectively. % The tuning of $\overline{\epsilon}_d$ can make the open system generate localized bound states or make the existed localized bound state(s) disappear. In another word, by increasing or decreasing the system on-site energy, some single-particle energy level(s) of the total system may be lifted~\cite{FH10} outside the continuous energy band and one or several bound states can be generated; or vice versa, energy level outside the energy band drops into the band and the corresponding bound state is transformed to the scattering state. To summarize, the impact of the static part of the driving on the decoherence may be constructive or destructive. Which case happens depends on the parameters.  

To illustrate the effect of the part $\Delta\epsilon_d(t)$ from the driving field more clearly, 
we would transform the total Hamiltonian to a new form. 
Formally, the time-independent part of the total Hamiltonian can be diagonalized. Because we focus on the decoherence 
dynamics controlling in this paper, without loss of generality, we may consider 
only one reservoir, namely a dot system coupled to a single lead  \cite{JTW13,House16} so that the index $\alpha$ in Eq.~(\ref{eq_H}) can be dropped.  
%making the total Hamiltonian in Eq.~\eqref{eq_Hp} can be expressed in terms of the field operators of its eigenmodes. 
The relation between the diagonalized eigenmodes and the original modes of the Hamiltonian reads~\cite{SH98,YLZ15,CZ21}  
\begin{subequations}\label{eq_transform}
\begin{align}
&c_{l_i}^\dag\!=\!\sqrt{Z_{i}} b^\dag+\sqrt{Z_{i}} \sum_{k}  \frac{V_{k}}{\epsilon_{l_i}-\epsilon_{k}} b_{k}^\dag\, ,\\
&c_{k}^\dag\!=\!b_{k}^\dag +V_{k} U(\epsilon_{k})\Big[b^\dag + \sum_{k'} \frac{V_{k'}}{\epsilon_{k}-\epsilon_{k'}+i 0} b_{k'}^\dag \Big]\,,
\end{align}
\end{subequations}
where $c_{l_i}^\dag$ and $c_{k}^\dag$ are the  creation operator of the $i$th localized bound state and scattering mode-$k$, respectively; $U(\epsilon_{k})\!=\!{1}/[{\epsilon_{k}\!-\!(\epsilon_s\!+\!\overline{\epsilon}_d)\!-\!\Sigma(\epsilon_{k})}]$ % $U(\epsilon_k)$
\iffalse
\begin{align}
U(\epsilon_{\alpha k})\!=\!\frac{1}{\epsilon_{\alpha k}-(\epsilon_s+\overline{\epsilon}_d)-\Sigma(\epsilon_{\alpha k})}\,,
\end{align} 
\fi 
is the retarded Green function in the energy domain. 
%standing for the self-energy correction~\cite{ZLX12}. 
In this new basis, the total Hamiltonian under driving can be written as
\begin{align}\label{eq_Htot}
H_{\rm tot}(t)\!=\!& \sum_i \epsilon_{l_i} c_{l_i}^\dag c_{l_i} \!+\! \sum_{k} \epsilon_{k} c_{k}^\dag c_{k} \!+\! \Delta\epsilon_d(t) \Big[ \sum_{i,i'} \lambda_{ii'} c_{l_i}^\dag c_{l_{i'}} \nonumber\\
&\!+\! \sum_{i, k} [\lambda_{i  k} c^\dag_{l_i} c_{ k} \!+ \! h.c.] \!+ \! \sum_{ k,k'} \lambda_{ k k'} c^\dag_{ k} c_{ k'} \Big]\,,
\end{align}
where the coefficients $\lambda_{i k}, \lambda_{ k k'}$  depend on the transformation in Eq.~\eqref{eq_transform}. 
One can see now that the driving $\Delta\epsilon_d(t)$ introduces the transitions among the localized bound states 
and the scattering modes. Naturally, \blk the form of $\Delta\epsilon_d(t)$ heavily influences the decoherence behavior of the 
open system. In the following two sections, we shall discuss these effects. One section is focused on the weak-driving case, while the other is focused on the case of strong-driving. It will be shown that although the picture of energy transfer works for both scenarios, 
the detailed dynamics can be quite different. \blk

\section{Weak driving field}
\label{sec3}

In the following, we shall illustrate that the behavior of $u(t,t_0)$ is mainly determined by the characteristics of $u_0(t,t_0)$ and the spectrum of the driving field for a weak driving field. 
\iffalse
In general, the $u_0(t,t_0)$ possesses the form~\cite{ZLX12}
\iffalse
\begin{align}
u_0(t,t_0)=
\end{align}
\fi 
\begin{align}\label{eq_u_conti}
u_{0}(t,t_0)%=\langle \{c(t), b^\dag(0)\}\rangle
=\int \frac{d \epsilon}{2\pi} D (\epsilon) e^{-i \epsilon (t-t_0)}\,,
\end{align}
with
\begin{align}\label{eq_dpec}
D(\epsilon)\!=\!2\pi\!\sum_i Z_{l_i} \delta(\epsilon-\epsilon_{l_i}) % \nonumber\\&
\!+\!\frac{J(\epsilon)}{[\epsilon\!-\!\overline{\epsilon}_d\!-\!\Delta(\epsilon)]^2 \!+\!J^2(\epsilon)/4}\,. 
\end{align}
In the formula, $D(\epsilon)$ is the spectrum of the system; $Z_{l_i}$ and $\epsilon_{l_i}$ stand for the amplitude and frequency of the $i$th localized bound state, respectively; $J(\epsilon)$ and $\Delta(\epsilon)$ are the spectral density and self-energy of the open system, respectively. 
\fi 

\begin{figure}[t]
\centering
\hspace{-1.1em}
	\includegraphics[width =8.5 cm]{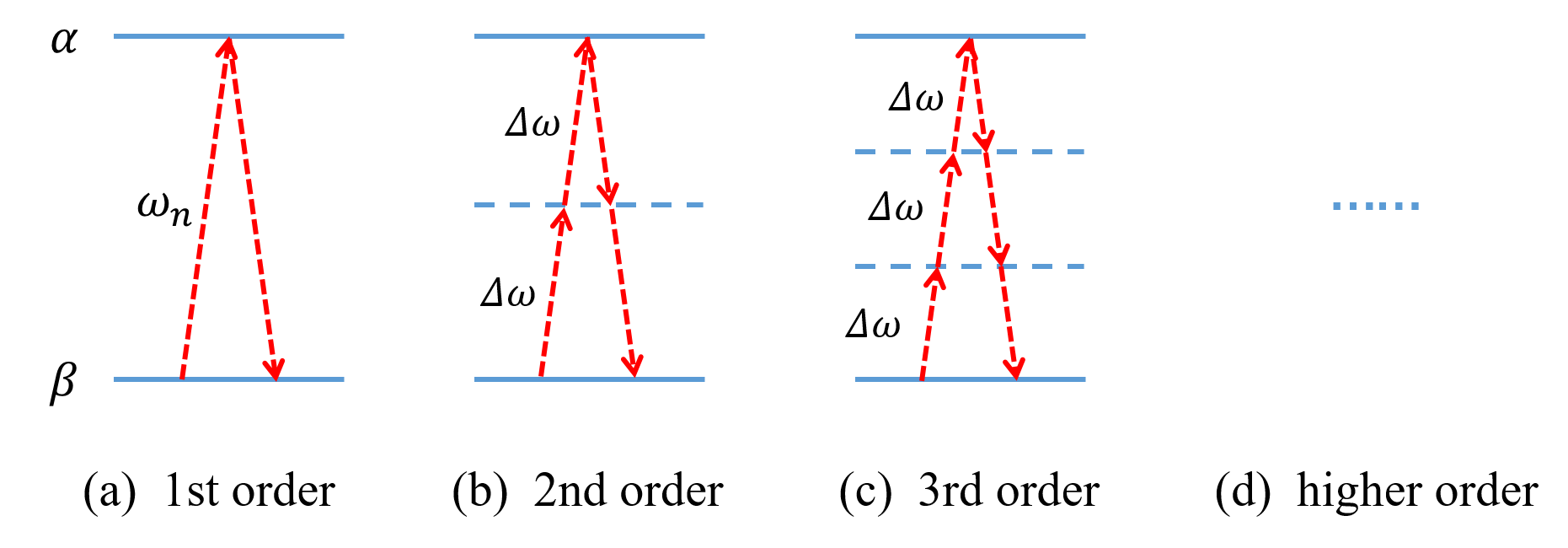}
	\caption{\small (Color online) The possible dominate processes when the driving is weak. 
	} \label{fig_transition}
\end{figure}

\begin{figure}
	\centering
	\subfigure[$A=0.5,T=1$]{\label{fig0a}
		\includegraphics[width=0.23\textwidth]{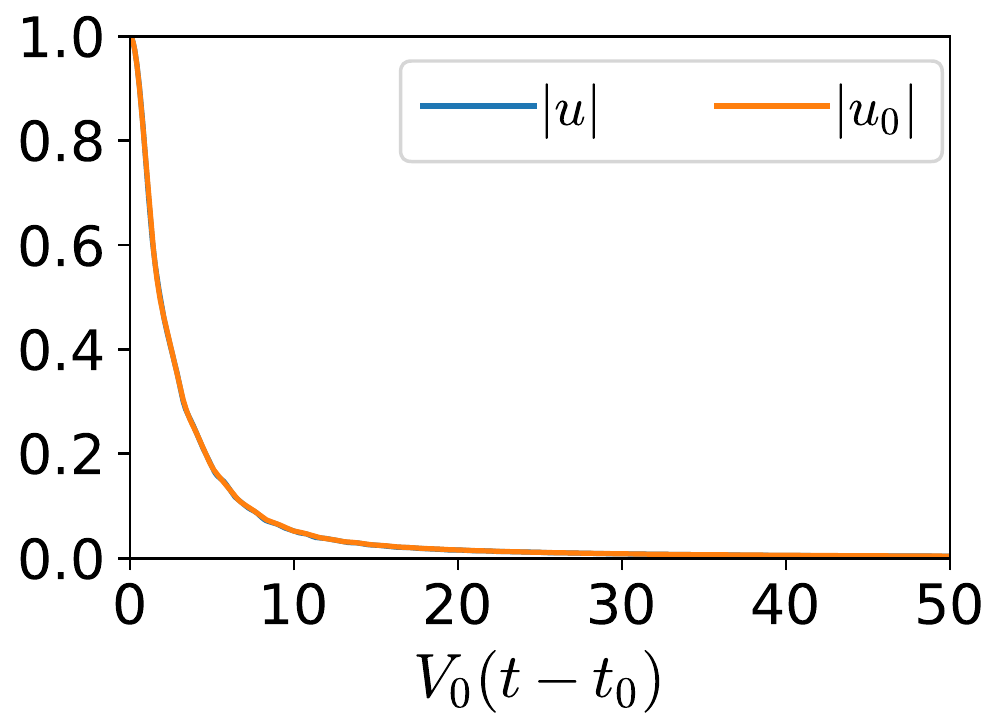}}\hspace{-0.8em}  
	\subfigure[$A=0.5,T=10$]{\label{fig0b}
		\includegraphics[width=0.23\textwidth]{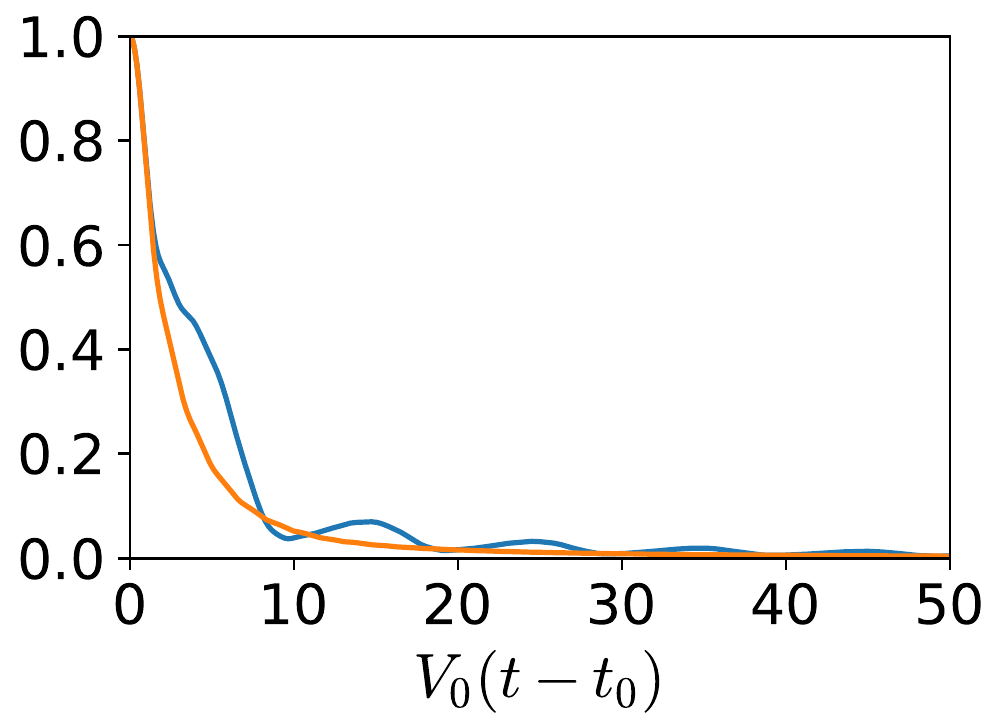}}\hspace{-0.8em}
	\caption
	{\small  (Color online) Plots of $|u_0(t,t_0)|$ and $|u(t,t_0)|$ for typical conditions without localized bound state in $u_0(t,t_0)$. The driving fields are both of the sine type. In the whole paper, we would keep using the same legends for the plots of $|u|$ and $|u_0|$. 
	}\label{fig0}
\end{figure}

The periodic driving field can always be decomposed as oscillations of different frequencies, i.e.,
\begin{align}
\Delta\epsilon_d(t)=\sum_{n=1}^{\infty} \big[ A_n \sin(\omega_n t) \!+\! B_n \cos(\omega_n t) \big]\,,
\end{align}
where $\omega_n\!=\! 2n\pi/T$, and $A_n$ and $B_n$ are the amplitudes of the oscillations. When the driving is weak, i.e., all the coefficients $A_n$ and $B_n$ are small, the effects of many terms in the time-dependent part
\begin{align}\label{eq_H_series}
\Delta\epsilon_d(t) \Big[ \sum_{i,i'} \lambda_{ii'} c_{l_i}^\dag c_{l_{i'}} \!+\! \sum_{i,k} [\lambda_{ik} c^\dag_{l_i} c_k \!+ \! h.c.] \!+ \! \sum_{k,k'} \lambda_{k k'} c^\dag_{k} c_{k'} \Big]
\end{align}
are negligible because of the energy conservation, which is similar to the dealing of the atom-photon interactions\blk. To be explicit, generally speaking, the terms involving with $e^{-i\omega_n t} c^\dag_\alpha c_\beta$ or its Hermitian conjugate, with the energy values satisfying $\hbar\omega_n=\epsilon_\alpha-\epsilon_\beta$ play the dominate role in the driving field. These terms represent the process of energy transfer between mode-$\alpha$ and $\beta$ through absorbing energy from or emitting energy to the driving field (see (a) of Fig.~\ref{fig_transition}). %The above is the first-order process. 
When the terms satisfying the above condition does not exist, the processes, which are of first order, are forbidden. In this case, higher-order process(es) may dominate. For instance, when the driving field $\epsilon_d(t)$ is of single frequency $\Delta\omega$ and the terms $e^{-i\Delta\omega t} c^\dag_\alpha c_\beta$ and $e^{i\Delta\omega t} c^\dag_\beta c_\alpha$ with $n\hbar\Delta\omega=\epsilon_\alpha-\epsilon_\beta$ exist in the series of Eq.~\eqref{eq_H_series}, higher order processes in (b)-(d) of Fig.~\ref{fig0} play the dominant role %. In this case, the processes $e^{-i\Delta\omega t} c^\dag_\alpha c_\beta$ and $e^{i\Delta\omega t} c^\dag_\beta c_\alpha$, with $n\hbar\Delta\omega=\epsilon_\alpha-\epsilon_\beta$,  dominate the time-dependent part of the Hamiltonian 
and determine the main properties of the decoherence dynamics. %(See (b)-(d) of Fig.~\ref{fig_transition}). 
These processes are the counterpart of the multi-photon process of in quantum optics~\cite{CDG98}.

In order to explain the idea of energy transfer introduced above, in the following,  we numerically discuss three different scenarios, i.e., there are no localized bound state, one localized bound state and multiple localized bound states in $u_0(t,t_0)$, respectively. Without loss of generality, we use a typical example that the system-environment interaction is described by the spectral density function~\cite{WLZX10}
\begin{align}
J(\epsilon)=
\begin{cases}
\eta^2 \sqrt{(2V_0)^2-(\epsilon-\epsilon_0)^2}\quad & %\text{if} \; 
|\epsilon-\epsilon_0|\leq 2 V_0\,, \\
0  &  \text{otherwise}.
\end{cases}
\end{align}
This spectral density exists in the model describing the impurity coupled to a chain of semi-infinite length, with $\epsilon_0$ standing for the on-site energy, $V_0$ quantifying the hopping rate and $\eta V_0$ characterizing the system-environment coupling strength. Due to the existence of two energy gaps, there may be zero, one or two localized bound states in $u_0(t,t_0)$~\cite{ZLX12}. In the following, we would set $t_0\!=\!0$, $\epsilon_0\!=\!0$ and $V_0\!=\!1$. If not specified, the pulse shape of the driving field $\Delta\epsilon_d(t)$ is set as the sine wave
\begin{align}\label{eq_sin}
\Delta\epsilon_d(t)\!=\!A \sin(2\pi t/T)\,,
\end{align}
\iffalse
\begin{align}
\phi(t)=-\frac{A}{\omega} \cos(\omega t)\,,
\end{align}
\begin{align}
\Delta\epsilon_d(t)\!=\!
\begin{cases}
\, A   &  t \in( t_0\!+\!nT, t_0\!+\!(n\!+\!\frac{1}{2})T] \,, \\
-A  &  t \in(t_0\!+\!(n\!+\!\frac{1}{2})T, t_0\!+\! (n\!+\!1)T] \,,
\end{cases}
\end{align}
\begin{align}
\phi(\tau)=\begin{cases}
\, A\operatorname{mod}({t-t_0,T})   &   t_0\!+\!nT \leq t \leq t_0\!+\!(n\!+\!\frac{1}{2})T  \\
-A ( ) &   t_0\!+\!(n\!+\!\frac{1}{2})T \leq t \leq  t_0\!+\! (n+1)T   \,,
\end{cases}  \,,
\end{align}
\fi 
with $A$ denoting the amplitude and $T$ standing for the fundamental period.  

% The Green function ${{u}}_0(t,t_0)$ is the solution of Eq.~\eqref{eq_u}, with the condition that $\epsilon_d(t)\!=\!\epsilon_0$. In another word, $u_0(t,t_0)$ is the Green function of the system with the static on-site energy.  

% \begin{figure}
% 	\centering
% 	\subfigure[$A=0.5,T=1$]{\label{fig0a}
% 	\hspace{-2.1em}
%     \includegraphics[width=0.55\textwidth]{fig0a}}
%     \\[-1.5ex]
% 	\subfigure[$A=0.5,T=10$]{\label{fig0b}
% 		\hspace{-2.1em}
%     \includegraphics[width=0.55\textwidth]{fig0b}} \\[-1.5ex]
% 	\subfigure[$A=3,T=2$]{\label{fig0c}
% 		\hspace{-2.1em}
%     \includegraphics[width=0.55\textwidth]{fig0c}}
% 	\caption
% 	{\small Plots of $|u_0(t)|$ and $|u(t)|$ for typical conditions without localized bound state in $u_0(t)$. In Fig.~\ref{fig0a} and~\ref{fig0b}, the driving field is of the sine type while in Fig.~\ref{fig0c}, it is the square wave.   
% 	}\label{fig0}
% \end{figure}

If there is no localized bound state in $u_0(t,t_0)$,  no time-dependent localized bound state can be generated by means of the periodic drive. In Fig.~\ref{fig0}, we set the parameters that $\eta\!=\! 0.8$ and $\overline{\epsilon}_d\!=\! 1$ to make that no localized bound state exists in $u_0(t,t_0)$. In Fig.~\ref{fig0a} and~\ref{fig0b}, the driving field is of the sine wave in Eq.~\eqref{eq_sin}, with the amplitudes the same while the fundamental periods varying much. When $T$ is short, the influence of the driving is quite limited such that the evolution of $u(t,t_0)$ is similar to that of $u_0(t,t_0)$ (see Fig.~\ref{fig0a}). When $T$ is long, the fundamental frequency $\Delta\omega$ is much smaller than the bandwidth of the spectral density. As a consequence, many more transitions  among the modes in the energy band (with respect to the short-$T$ case) happen, making the driving field impose a more influential effect on the open system (see Fig.~\ref{fig0b}). That is, it would induce oscillations around $u_0(t,t_0)$. However, even though it indicates a strong memory effect, no localized bound state can be generated. 
%Actually, it is a universal conclusion (for this case). 
Because no instantaneous localized bound state exists when the driving is weak, the instantaneous energy band of the total system is always continuous, which makes the particle in the system inevitably dissipate into the environment.

% Fig.~\ref{fig0c} demonstrates such a scenario by setting the driving field as the square wave. The amplitude is so large that during $nT$ to $(n\!+\!\frac{1}{2})T$, the instantaneous localized bound state is on one side of the energy band while from $(n\!+\!\frac{1}{2})T$ to $(n\!+\!1)T$, it is on the other side. As $\Delta\epsilon_d(t)$ changes sign, the previous instantaneous localized bound state is no longer localized, making the particle in it dissipates to the environment.  

\begin{figure}
	\centering
	\subfigure[$A=0.5,T=1.25$]{\label{fig1a}
		%\hspace{-2.1em}
		\includegraphics[height=0.13\textwidth]{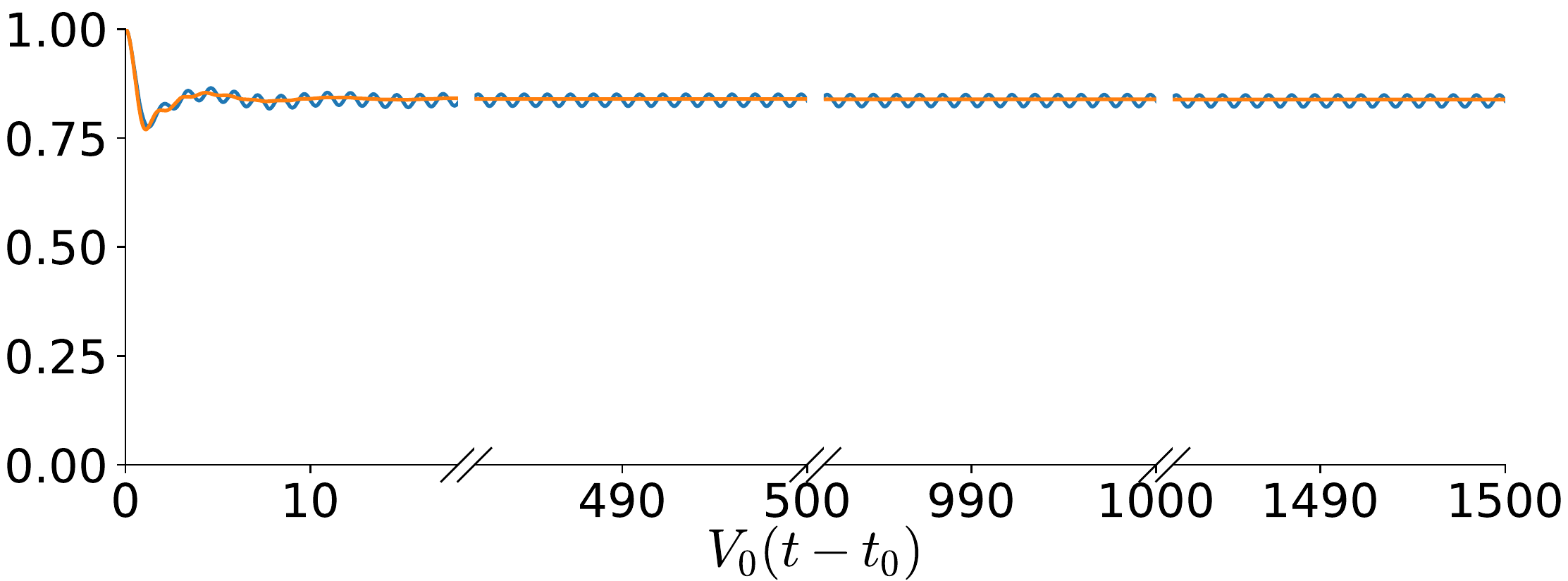}
		\hspace{-0.5em}  
		\includegraphics[height=0.13\textwidth]{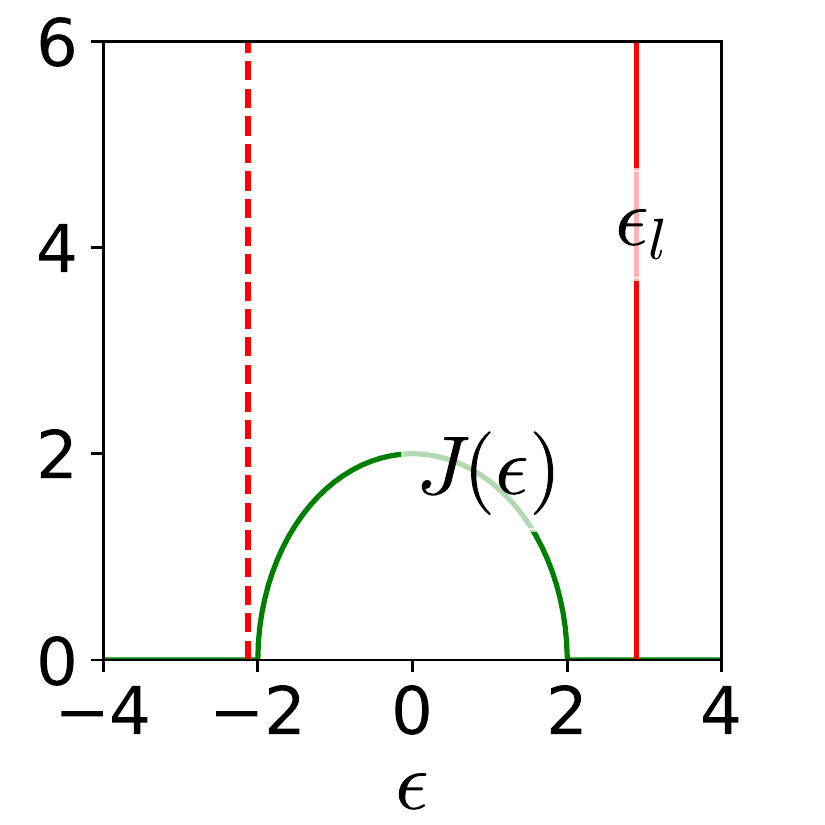}}
		\\[-1ex]
	\subfigure[$A=0.5,T=1.32$]{\label{fig1b}
		%\hspace{-2.1em}
		\includegraphics[height=0.13\textwidth]{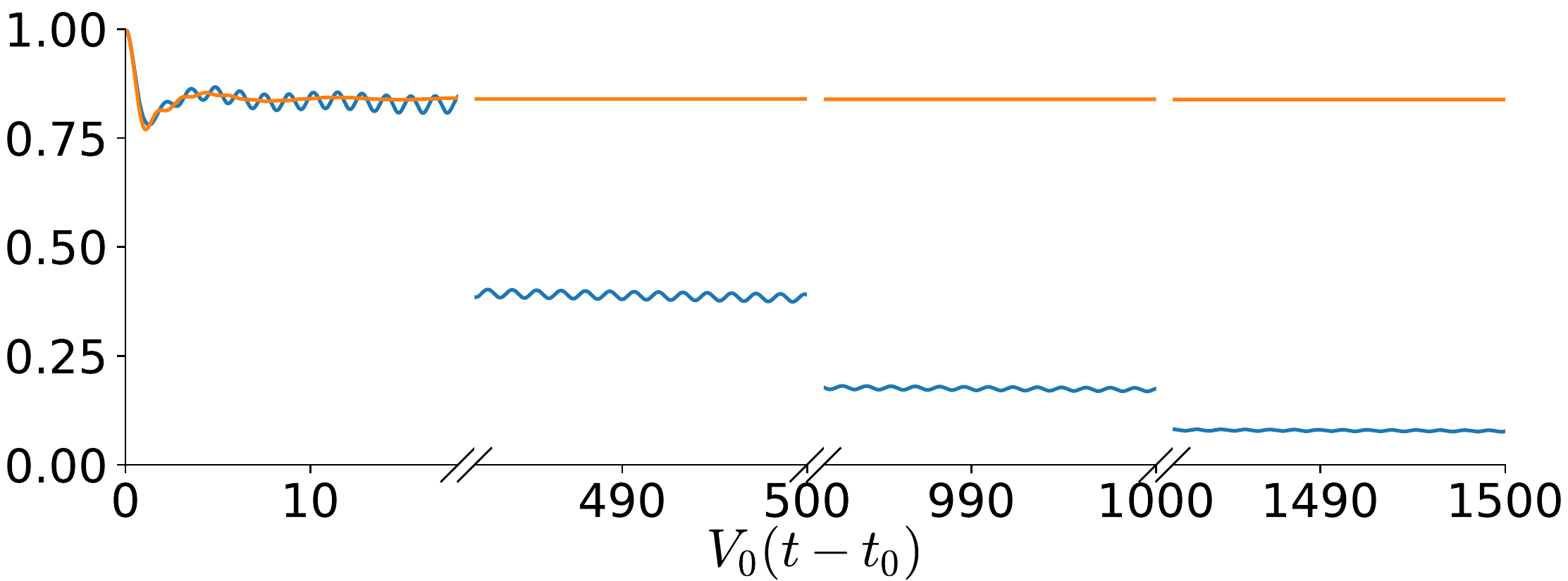}
		\hspace{-0.5em}  
		\includegraphics[height=0.13\textwidth]{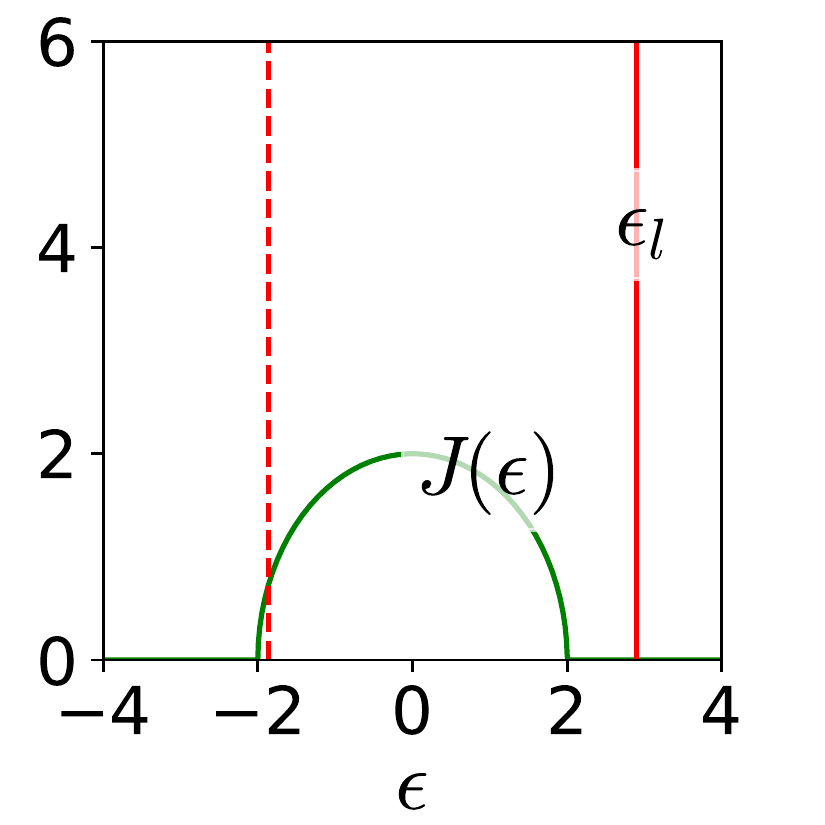}}
		\\[-1ex]
	\subfigure[$A=0.5,T=10$]{\label{fig1c}
		%\hspace{-2.1em}
		\includegraphics[height=0.13\textwidth]{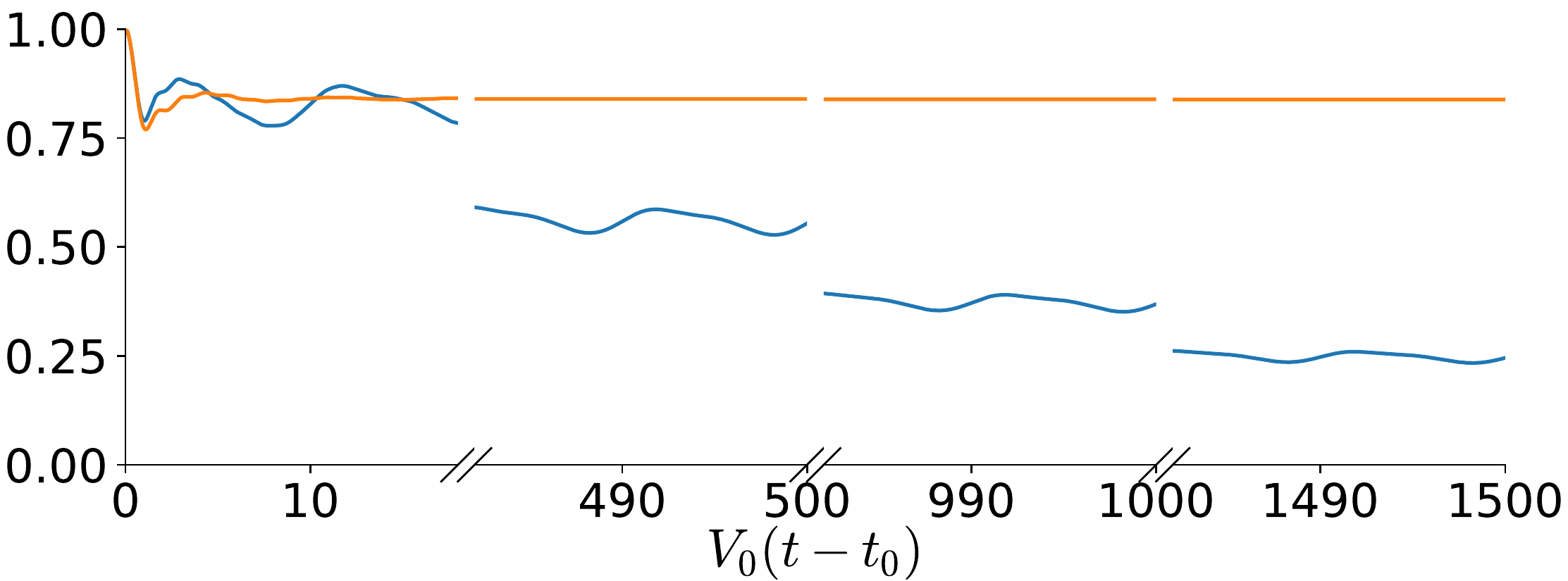}
		\hspace{-0.5em}  
		\includegraphics[height=0.13\textwidth]{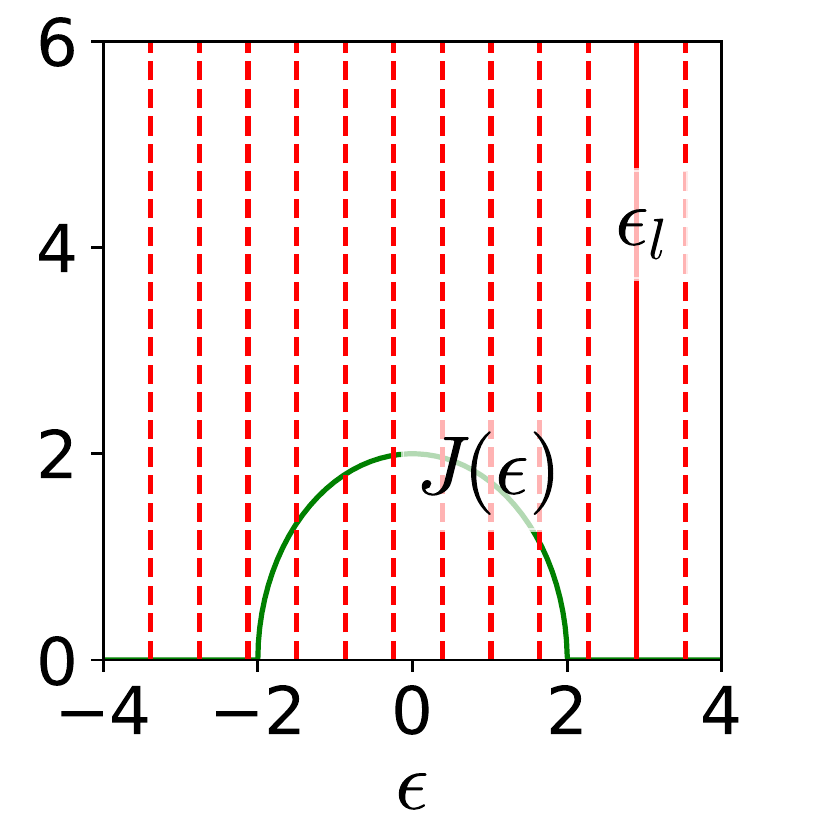}}
		\\[-1ex]
	\subfigure[$A=0.5,T=10$]{\label{fig1d}
		%\hspace{-2.1em}
		\includegraphics[height=0.13\textwidth]{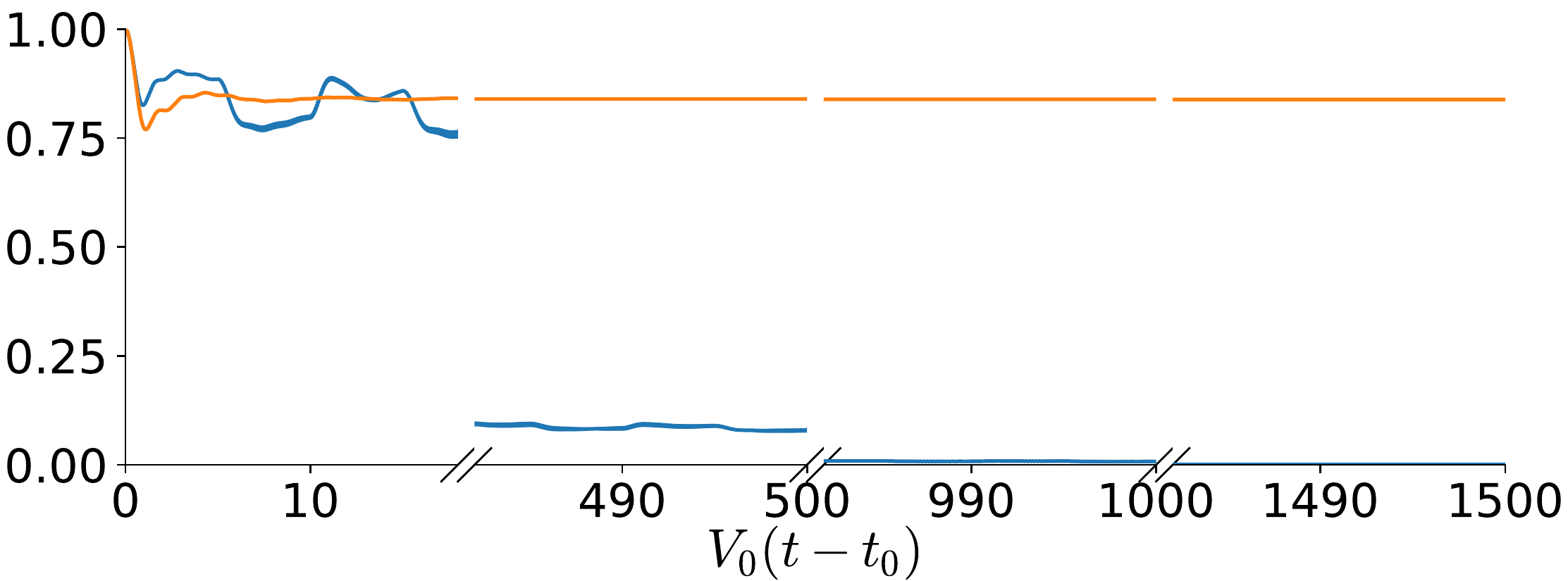}
		\hspace{-0.5em}  
		\includegraphics[height=0.13\textwidth]{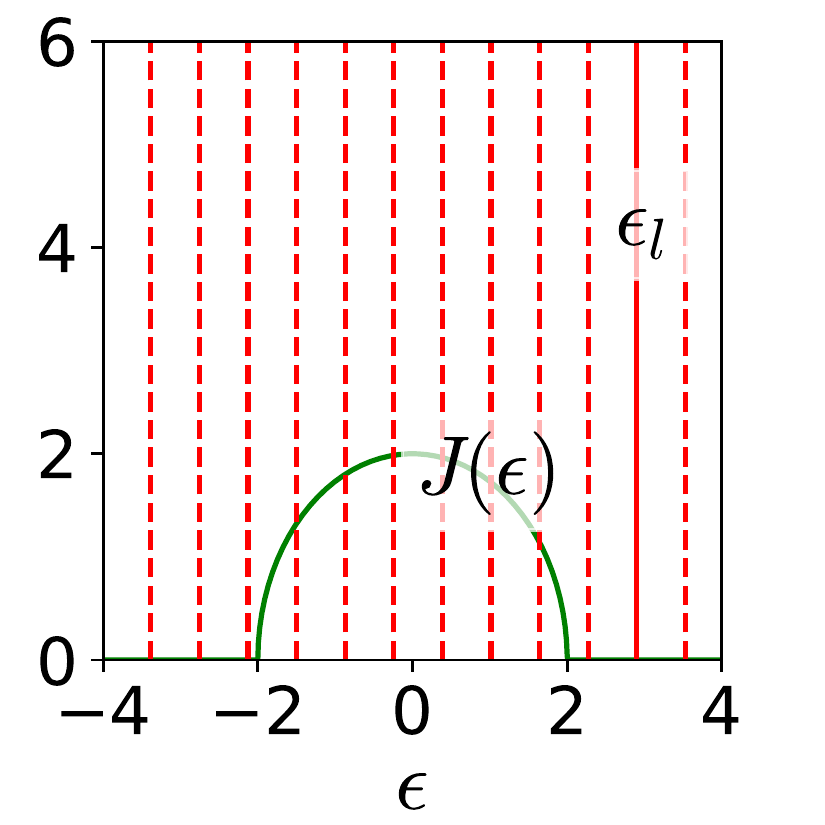}} \\[-1ex]
% 	\subfigure[$A=0.5,T=10$]{\label{fig1e}
% 		\includegraphics[width=0.8\textwidth]{fig1e}}\hspace{-0.8em}  
% 	\subfigure[$A=0.5,T=10$]{\label{fig1f}
% 		\includegraphics[width=0.19\textwidth]{fig1f}}\hspace{-0.8em} \\[-1ex]
	\caption
	{\small  (Color online) Plots of $|u_0(t,t_0)|$ and $|u(t,t_0)|$ for typical conditions with single localized bound state in $u_0(t,t_0)$ in the case of weak driving field. The left panels demonstrate the evolution of $|u_0(t,t_0)|$ and $|u(t,t_0)|$, while the right ones are the illustration of the spectra, with the green line standing for the plot of $J(\epsilon)$, the red solid line and the red dashed lines being the plots of $\epsilon=\epsilon_l$ and $\epsilon=\epsilon_l+n\hbar\Delta\omega$, respectively. In (a)-(c), the driving fields are of sine wave, which demonstrate the typical case that the spectral density functions has no overlap with the series $\epsilon_l\!+\!n\hbar\Delta\omega$, overlaps with $\epsilon_l-\hbar \Delta\omega$, and overlaps with higher-order frequencies. In (d), the driving field is of the square wave, with the amplitude $A$ and the fundamental period $T$ the same as those in (c). 
	}\label{fig1}
\end{figure}

In Fig.~\ref{fig1}, we set the parameters that $\eta\!=\!1$ and $\overline{\epsilon}_d\!=\!2.5$, which is made to guarantee the existence of one single localized bound state in $u_0(t,t_0)$ (see the solid red lines denoting the localized bound state energy in the right panels). The parameters $A$ and $T$ are shown below the corresponding subfigures. In Fig.~\ref{fig1a}-\ref{fig1c}, the pulse shapes are sinusoidal (see Eq.~\eqref{eq_sin}) while in Fig.~\ref{fig1d}, it is the square wave (see Eq.~\eqref{eq_square}). From Fig.~\ref{fig1a}-\ref{fig1c}, the frequency of  $\Delta\epsilon(t)$ increases. In Fig.~\ref{fig1a}, the frequencies $\epsilon_l\!+\!n\hbar\Delta\omega$ (where $n$ is an integer) has no overlap with the spectral density function $J(\epsilon)$, thus no transition between the localized bound state and the continuous band can happen. As a result, the localized bound state cannot be destroyed by the periodic field and $u(t,t_0)$ does not dissipate to zero. In Fig.~\ref{fig1b}, $\epsilon_l-\hbar\Delta\omega$ is in the range of $J(\epsilon)$, making that particle in the localized bound state can release an energy quantum $\hbar\Delta\omega$ and jump to the continuous energy band. As a result, the driving field generates a dissipation channel of the original localized bound state, making $u(t,t_0)$ finally decay to zero. In Fig.~\ref{fig1c}, $\epsilon_l-\hbar\Delta\omega$ is not in the range of $J(\epsilon)$ while $\epsilon_l-2 \hbar\Delta\omega$, $\cdots$, $\epsilon_l-7 \hbar\Delta\omega$ is in it. Consequently, the original localized bound state would release at least two energy quanta to make the dissipation happen. This should involve with several high-order processes, so that the Green's function $u(t,t_0)$ would finally decay to zero but with a smaller rate with respect to that in Fig.~\ref{fig1b}. % The behavior difference of the $u(t,t_0)$'s in Fig.~\ref{fig1c} and ~\ref{fig1b} is  that the dissipation mechanisms are different. That is, Fig.~\ref{fig1b} is a first-order process and Fig.~\ref{fig1c} is composed of several higher-order processes, which is similar to the multi-photon processes widely studied in quantum optics~\cite{CDG98}. 
The viewpoint can be further supported by comparing Fig.~\ref{fig1d} with Fig.~\ref{fig1c}. In these two figures, the parameters $A$ and $T$ are the same, except that the waveforms are different, i.e., in Fig.~\ref{fig1d}, $\Delta\epsilon_d(t)$ is the square wave
\begin{align}\label{eq_square}
\Delta\epsilon_d(t)\!=\!
\begin{cases}
\, A   &  nT \leq t \leq (n+\frac{1}{2})T \\
-A  &  (n+\frac{1}{2})T \leq t \leq   (n\!+\!1)T \,.
\end{cases}
\end{align}
Even though the fundamental frequencies are the same, because $\Delta{\epsilon}_d(t)$ in Fig.~\ref{fig1d} naturally possesses frequency components of $2\Delta\omega$, $3\Delta\omega$, $\cdots$. The pulses can induce dissipation with the first-order process of (a) in Fig.~\ref{fig_transition}. The consequence is that the decay rate is much faster than that in Fig.~\ref{fig1c}.

\begin{figure}
	\centering
	\subfigure[$A=0.5,T=1.25$]{\label{fig2a}
		%\hspace{-2.1em}
		\includegraphics[height=0.13\textwidth]{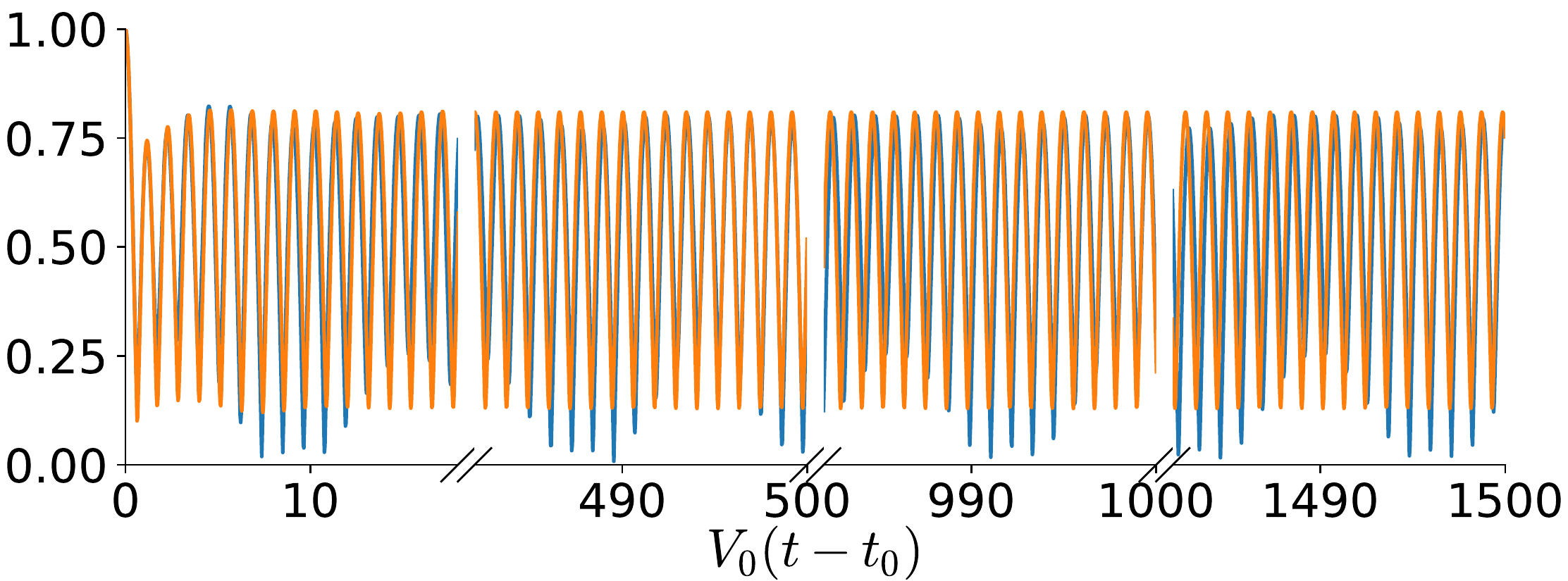}
		\hspace{-0.5em}  
		\includegraphics[height=0.13\textwidth]{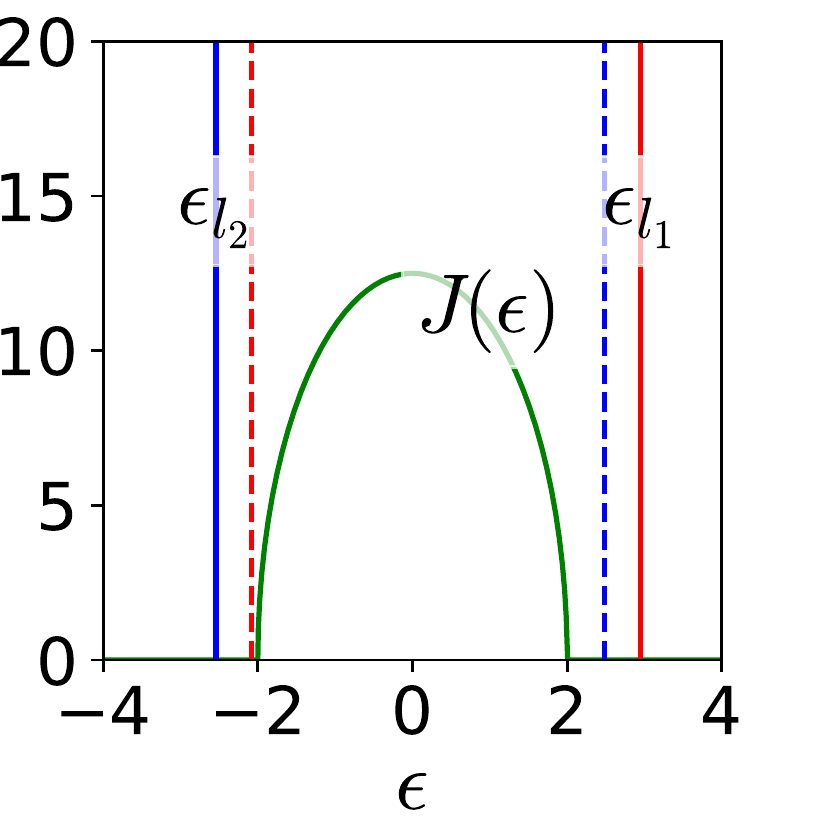}}%\hspace{-0.8em} 
		\\[-1ex]
	\subfigure[$A=0.5,T=1.32$]{\label{fig2b}
		%\hspace{-2.1em}
		\includegraphics[height=0.13\textwidth]{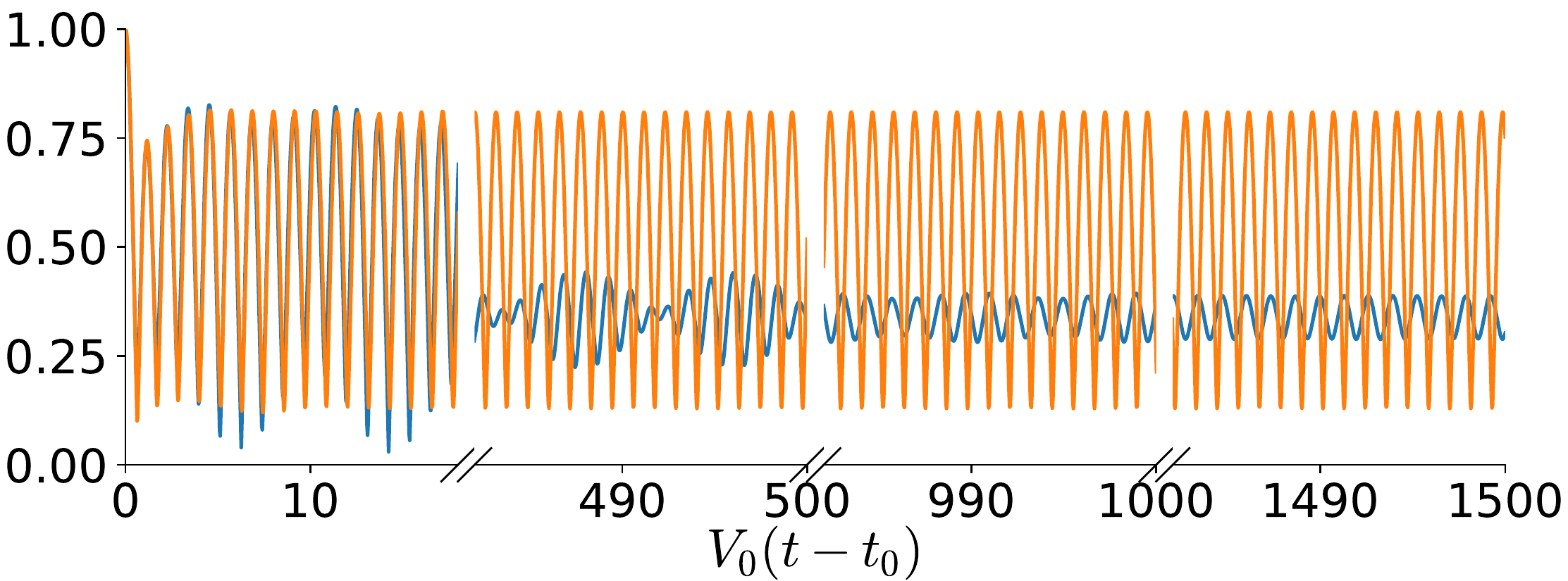}
		\hspace{-0.5em}  
		\includegraphics[height=0.13\textwidth]{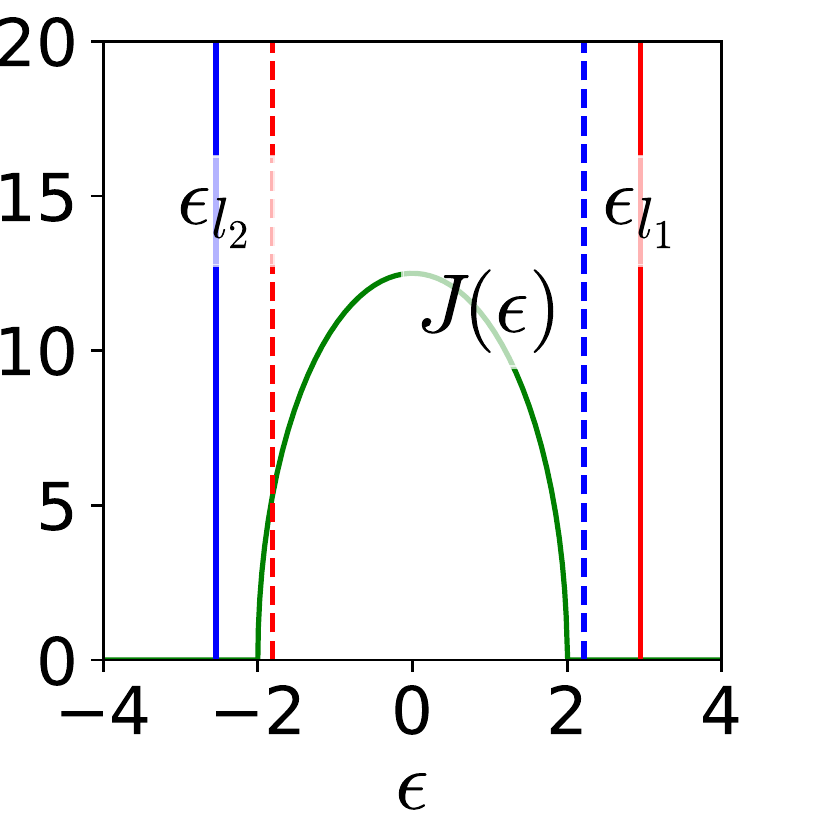}}%\hspace{-0.8em} 
		\\[-1ex]
	\subfigure[$A=0.5,T=2$]{\label{fig2c}
		%\hspace{-2.1em}
		\includegraphics[height=0.13\textwidth]{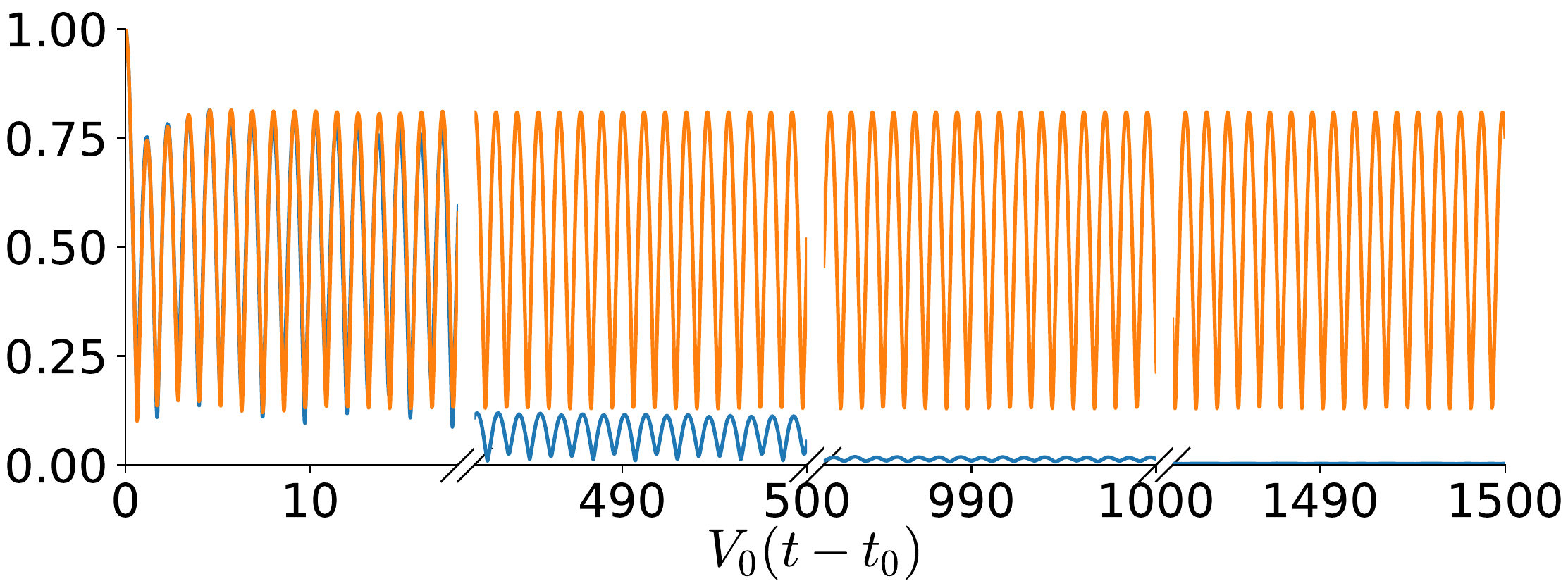}
		\hspace{-0.5em}  
		\includegraphics[height=0.13\textwidth]{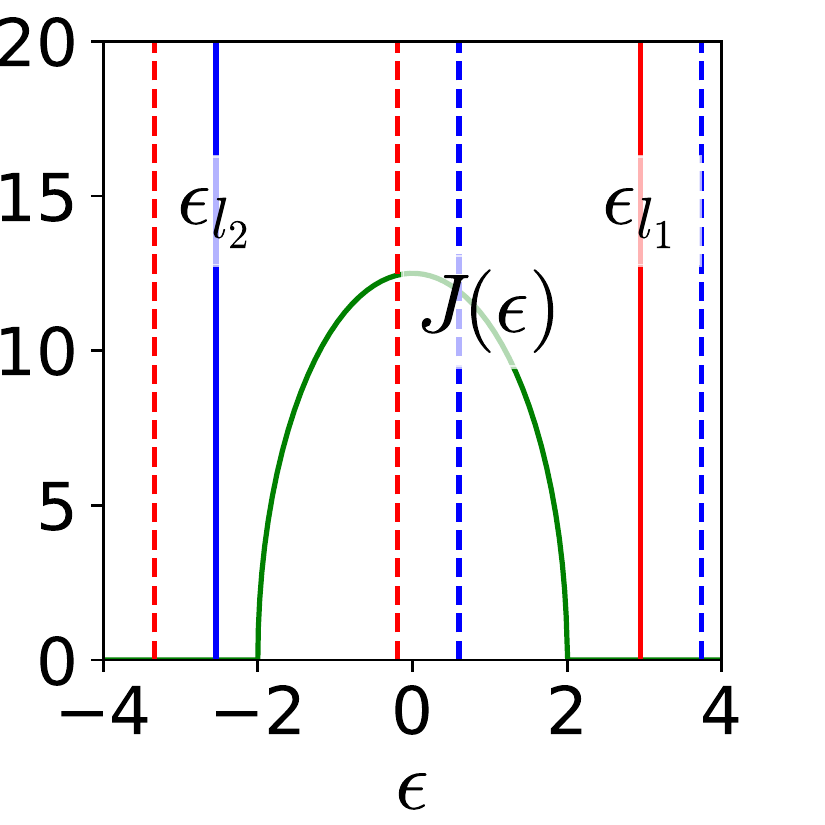}}%\hspace{-0.8em} 
		\\[-1ex]
	\subfigure[$A=0.5,T=2\pi\hbar/(\epsilon_{l_1}-\epsilon_{l_2})$]{\label{fig2d}
		%\hspace{-2.1em}
		\includegraphics[height=0.13\textwidth]{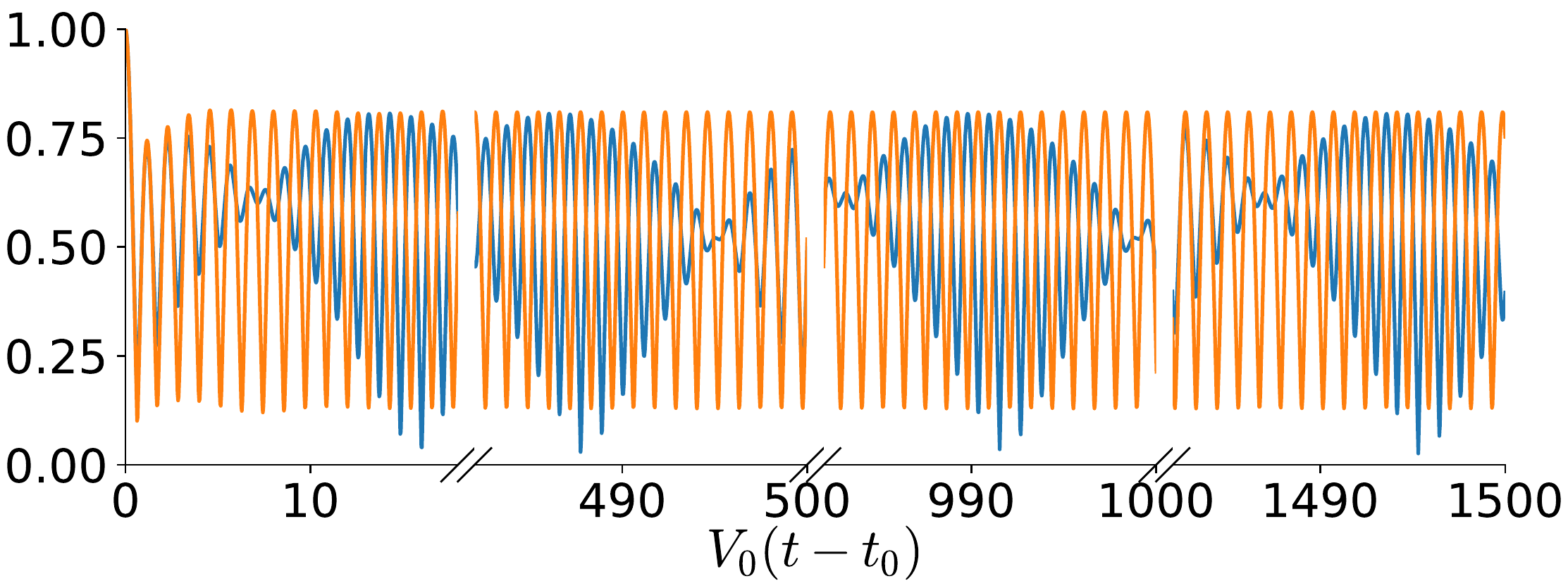}
		\hspace{-0.5em}  
		\includegraphics[height=0.13\textwidth]{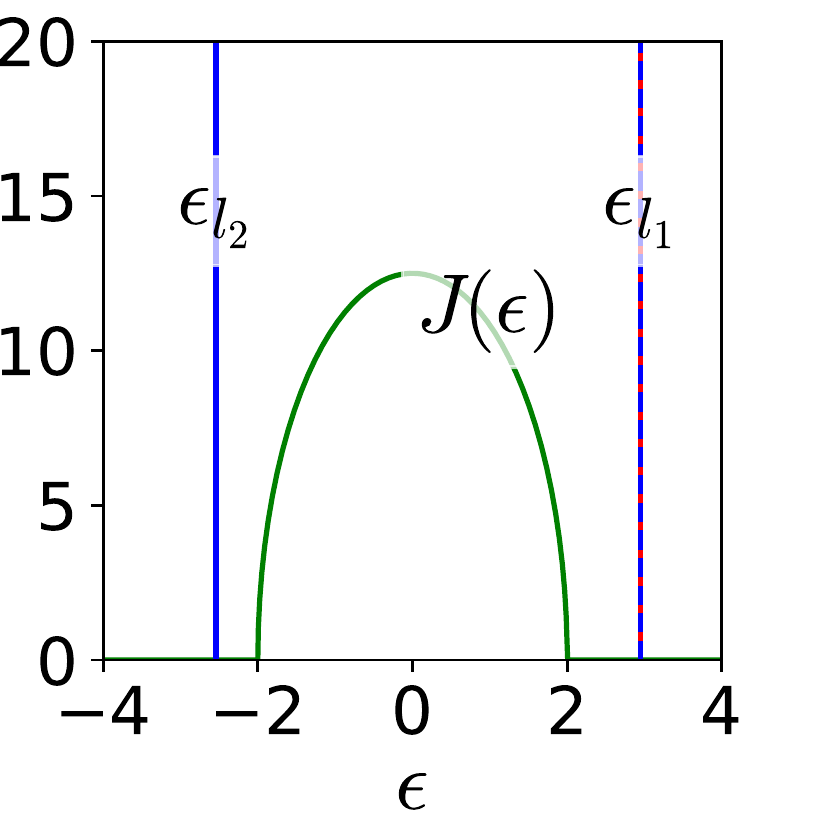}}%\hspace{-0.8em} 
		\\[-1ex]
	\caption
	{\small  (Color online) Plots of $|u_0(t,t_0)|$ and $|u(t,t_0)|$ for typical conditions with multiple localized bound states in $u_0(t,t_0)$. The conventions of the plots are the same as those in Fig.~\ref{fig1}, except that there are two localized bound states in this case and in the right panels the series of plots $\epsilon=\epsilon_{l_1}\!+\!n\hbar\Delta\omega$ and $\epsilon=\epsilon_{l_2}\!+\!n\hbar\Delta\omega$ are colored as red and blue, respectively. From (a) to (c), the fundamental period increases such that in these cases, no frequency comb $\epsilon_{l_i}\!+\!n\hbar\Delta\omega$, a single frequency comb, and both frequency combs overlap(s) with $J(\epsilon)$. In (d), the fundamental frequency is chosen so that $\Delta\omega$ is exactly the energy difference of the two localized bound states.  % In Figs.~\ref{fig2c} and \ref{fig2d}, the upper one involves the first-order process while the bottom one involves high-order processes, making the dissipation rate so different. 
	}\label{fig2}
\end{figure}

When there are multiple localized bound states in $u_0(t,t_0)$, the phenomena would be more abundant. The dissipation of the localized bound states possesses the property of independence. % which is also the result of the energy transition picture in Fig.~\ref{fig_transition}. 
That is, if the localized bound state $l_i$ exists such that $\epsilon_{l_i}\!+\!n\Delta\omega$ overlaps with the spectral density function, then it would dissipate and if not, it would not dissipate. \blk In Fig.~\ref{fig2}, we  %demonstrate the phenomena with the numerical method. 
set the parameters $\overline{\epsilon}_d\!=\!0.5$ and $\eta\!=\!2.5$ to generate two localized bound states in $u_0(t,t_0)$, with the energy configuration shown in the right panels. The waveforms of the driving field are all chosen as sinusoidal with the amplitudes and fundamental periods demonstrated in the captions. In Fig.~\ref{fig2a}, there is no overlap for either localized bound state. Therefore, the original localized bound states cannot exchange energy with the continuous band, making them both survive. In Fig.~\ref{fig2b}, the spectrum series of $l_1$ overlaps with $J(\epsilon)$, therefore, only $l_2$ would survive in the long-time limit of $u(t,t_0)$. In Fig.~\ref{fig2c}, the overlap happens for both localized bound states. As a consequence, both localized bound states possess the dissipation channel and $u(t,t_0)$ would finally decay to zero. One special case of Fig.~\ref{fig2a}, in which the special means that the fundamental frequency of the driving exactly matches the energy difference of the localized bound states, is plotted in Fig.~\ref{fig2d}. Both localized bound states survive as those in Fig.~\ref{fig2a}, except for the more obvious beating character revealed in the evolution of $u(t,t_0)$.

% The only difference is that Fig.~\ref{fig2c} corresponds to the first-order process while Fig.~\ref{fig2d} is involved with higher-order processes, making the decay rates very different.  

In Section~\ref{sec2}, we have roughly introduced the effect of the static component of the driving field that it can be either constructive or destructive in controlling the decoherence. In this section, we show that the zero-averaged weak driving field does not impose the constructive effect in controlling decoherence. More specifically, it cannot generate new localized bound states but can destroy the originally existed localized bound states. The destroying process is mediated by exchanging energy between the localized bound state and the continuous energy band. Without the driving field, such a process cannot happen because of the energy conservation law. While with it, particle in the localized bound state can absorb or release energy from the field and transit to the state in the continuous band. If this process can happen, then the localized bound state would be destroyed. If the driving field cannot make it happen, then the system would have periodic localized bound state(s).

\iffalse
\begin{align}
\Delta\epsilon_d(t)\!=\!A \sin[\omega (t-t_0)]\,,
\end{align}

\begin{align}
\phi(\tau)=-\frac{A}{\omega} \cos[\omega(\tau-t_0)]\,,
\end{align}
\fi

\iffalse
\begin{align}
\phi(t)=\begin{cases}
\, A\operatorname{mod}({t,T})   &   nT < t \leq (n\!+\!\frac{1}{2})T  \\
A [T-\operatorname{mod}({t,T})] &   (n\!+\!\frac{1}{2})T \leq t < (n\!+\!1)T   \,,
\end{cases} 
\end{align}
and $\phi(nT)=0$. 
\fi

\section{Strong driving field}
\label{sec4}

\begin{figure}
	\centering
	\subfigure[$A=5,T=1.25$]{\label{fig3a}
		\includegraphics[height=0.13\textwidth]{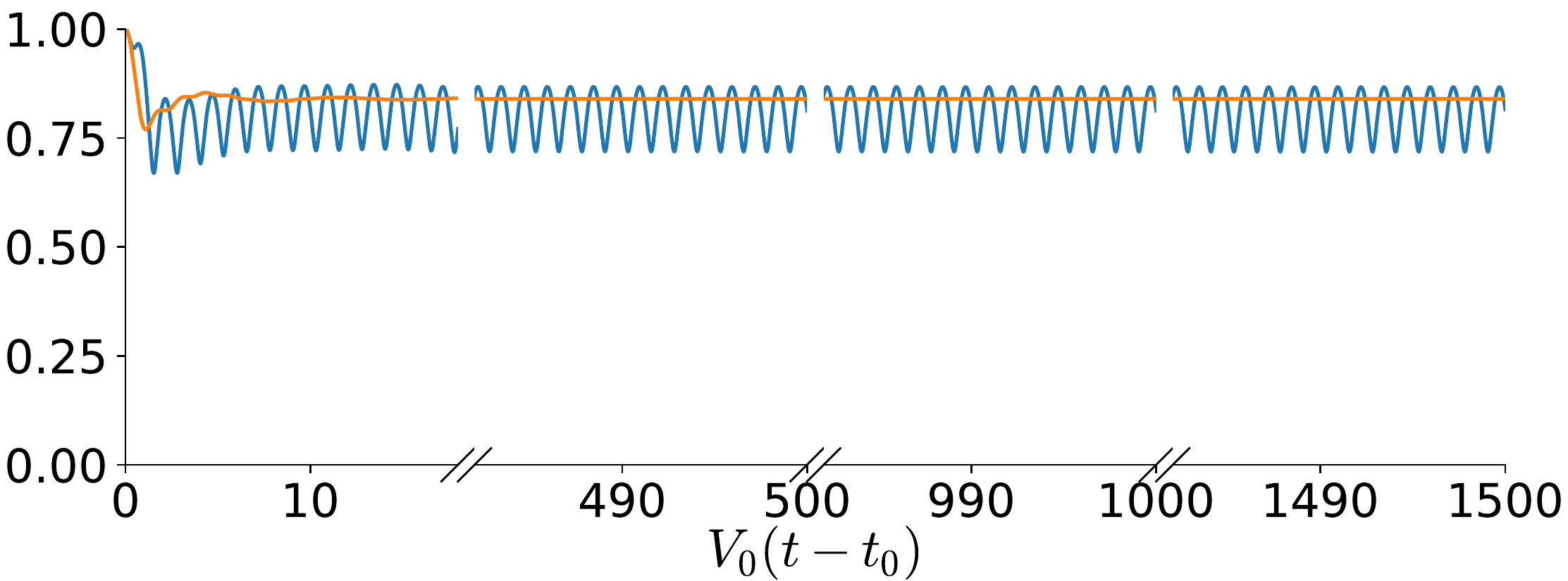}
		\hspace{-0.5em}  
		\includegraphics[height=0.13\textwidth]{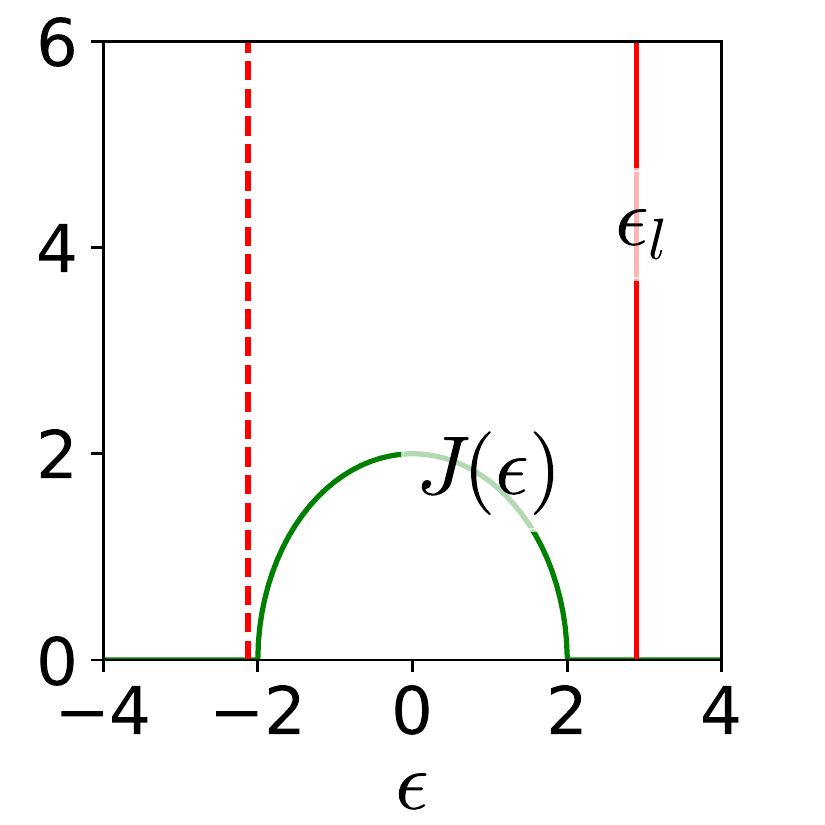}}
		\\[-1ex]
	\subfigure[$A=5,T=1.32$]{\label{fig3b}
		\includegraphics[height=0.13\textwidth]{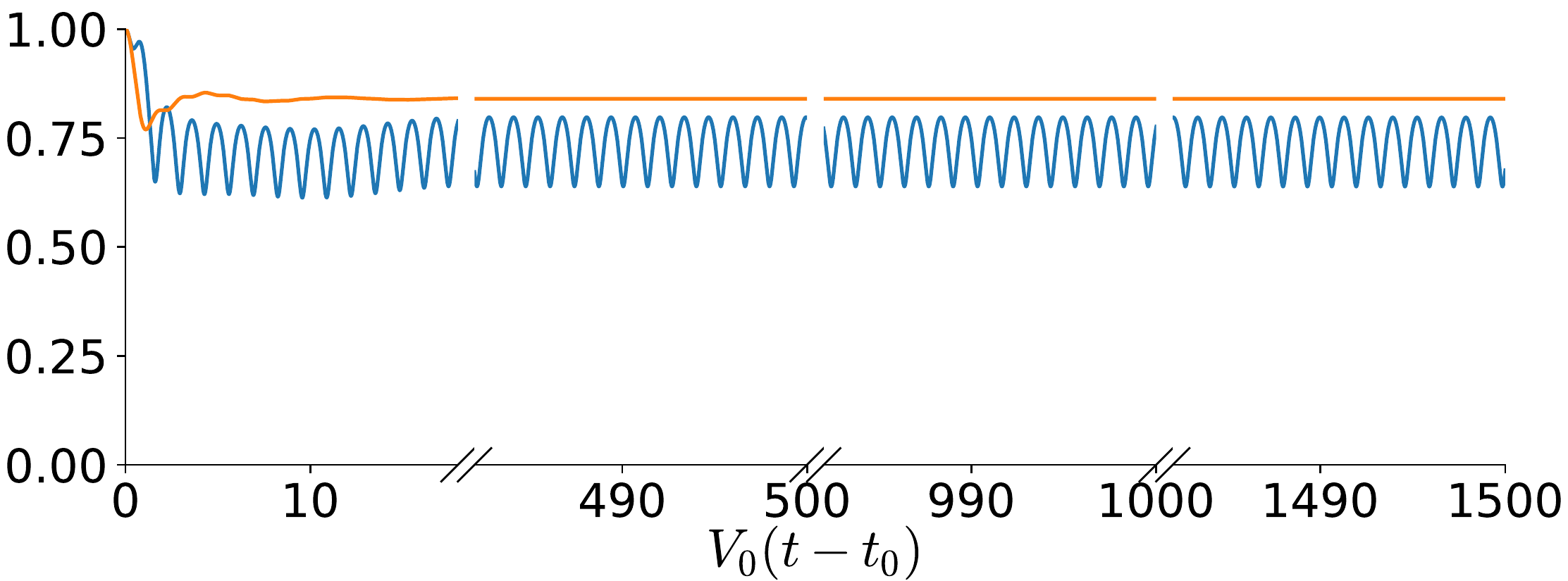}
		\hspace{-0.5em}  
		\includegraphics[height=0.13\textwidth]{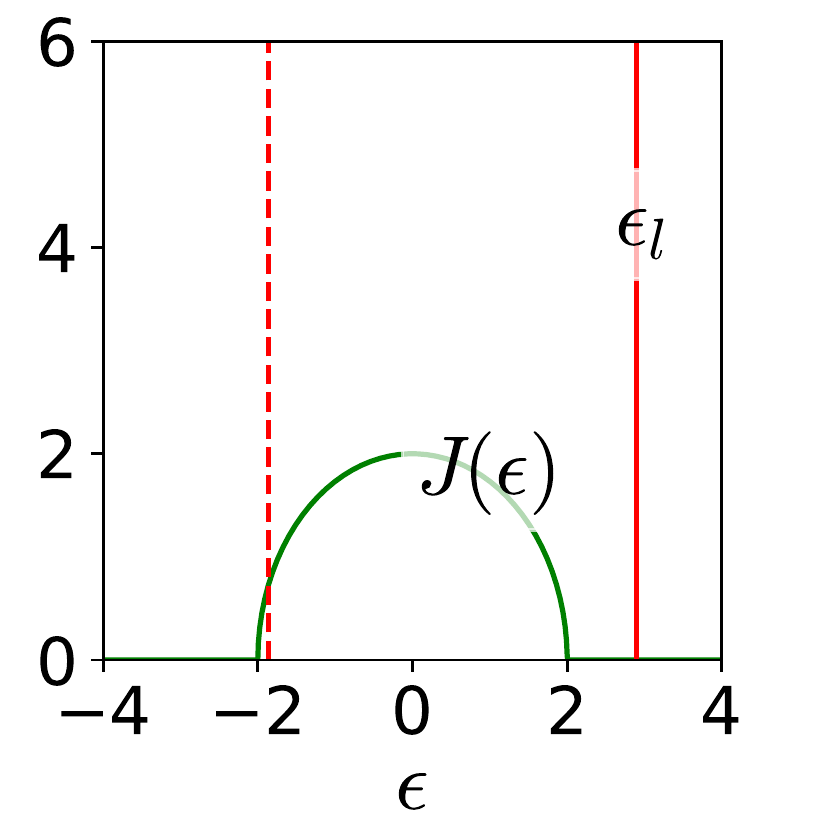}}
		\\[-1ex]
	\subfigure[$A=5,T=1.405$]{\label{fig3c}
		\includegraphics[height=0.13\textwidth]{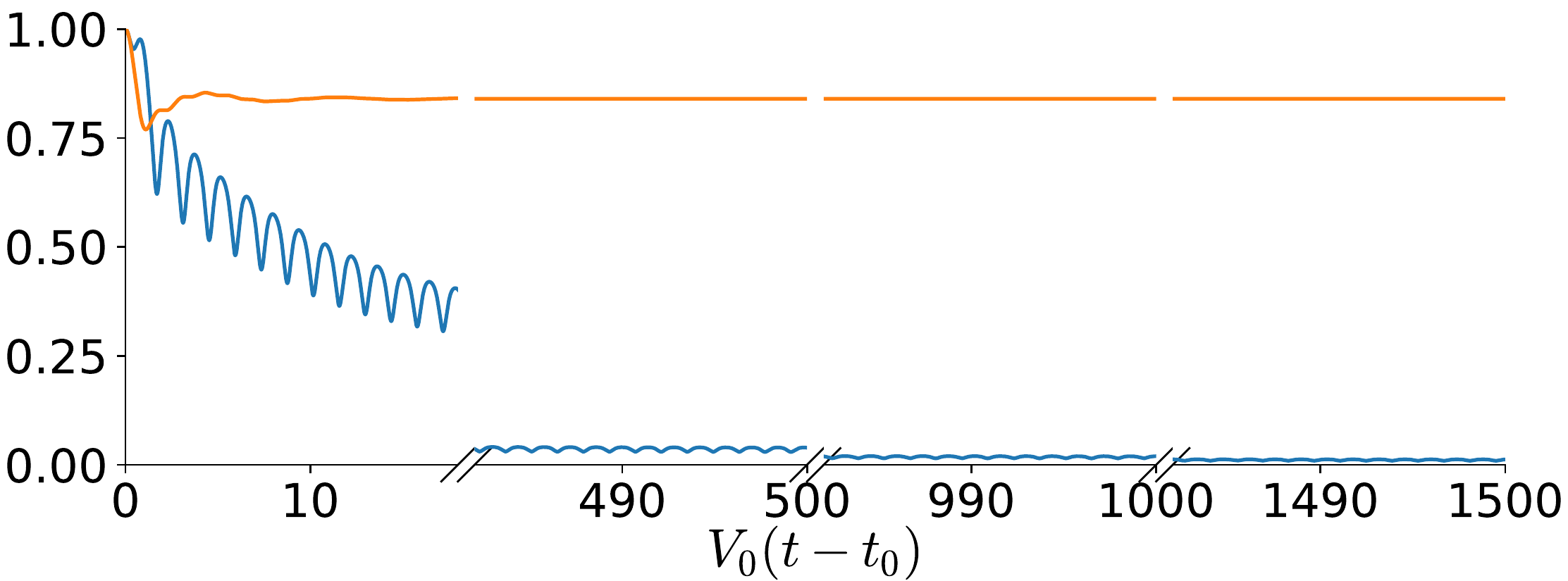}
		\hspace{-0.5em}  
		\includegraphics[height=0.13\textwidth]{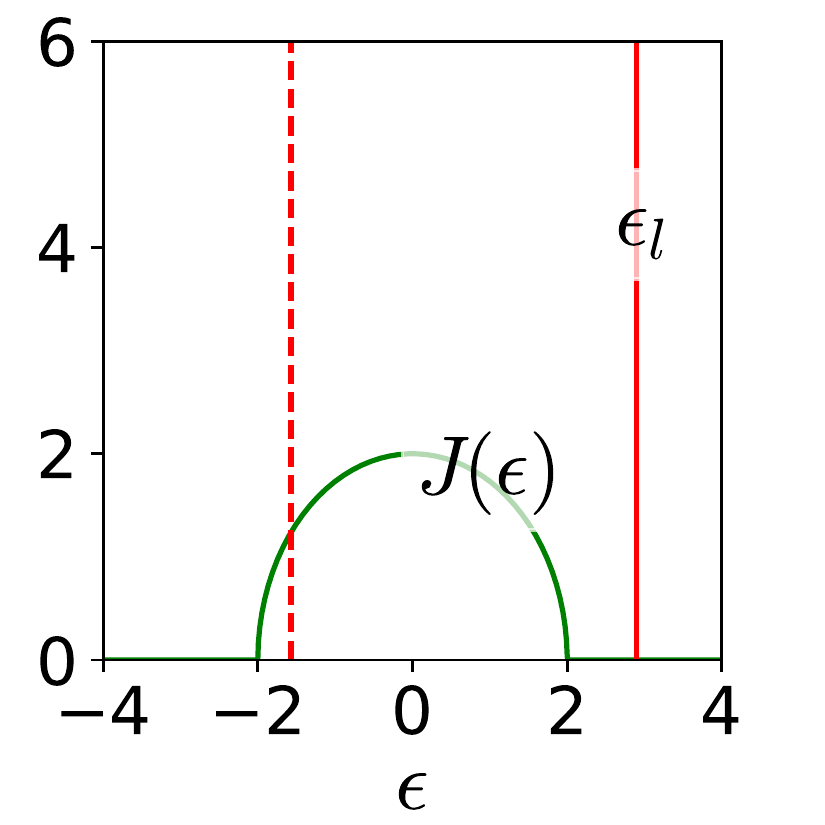}}
		\\[-1ex]
	\caption
	{\small  (Color online) Plots of $|u_0(t,t_0)|$ and $|u(t,t_0)|$ for typical conditions with single localized bound state in $u_0(t,t_0)$ in the case of strong driving field. The driving fields are of sine wave, which demonstrate the typical case that the energy value $\epsilon_l\!-\!\hbar\Delta\omega$ is around the band edge of the environment. 
	}\label{fig3}
\end{figure}

When the driving field is strong, the dynamical picture of energy transfer in the weak field case should be modified. This can be clearly revealed by the phenomena shown in Fig.~\ref{fig3}. To demonstrate clearly the difference , we adopt the parameter settings that $\eta\!=\!1$ and $\overline{\epsilon}_d\!=\!2.5$ (which are the same as those in Fig.~\ref{fig1}) such that a single localized bound state exists when the driving $\Delta\epsilon_d(t)$ is absent.

In Fig.~\ref{fig3}, the fields are all in the sine shape with amplitude set as $A\!=\!5$, being much larger than those in Fig.~\ref{fig1}. The fundamental frequencies in the top two subfigures are the same as those in Fig.~\ref{fig1}, respectively. Note that for $T\!=\!1.25$, the energy values $\epsilon_l+n\hbar\Delta\omega$ do not overlap with the spectral density function $J(\epsilon)$ (See Fig.~\ref{fig3b}). Although the modulation to the localized bound state is much larger than that in the weak driving case, the localized bound state survives. The comparing between Figs.~\ref{fig3b} and ~\ref{fig1b} reveals the essential difference between weak driving and strong driving, which shows that the picture of the energy transfer should be modified in the case of a strong driving field. Note that for the strong driving, even though the energy value $\epsilon_l-\hbar\Delta\omega$ is in the environmental energy band, the localized bound state can still survive. When the fundamental frequency increases further such that $\epsilon_l-\hbar\Delta\omega$ is not so close to the band edge (See Fig.~\ref{fig3c}), the localized bound state can be destroyed completely.

The reason why the strong driving case breaks the picture in the previous section lies in the energy correction. Actually, when a sinusoidal driving field is imposed on the system, the energy level corresponding to the localized bound state would emit and reabsorb the energy quanta, making its energy shift~\cite{CDG98}. When the driving is strong, the shift is not small and must be taken into account when dealing with the system decoherence dynamics. As a consequence, the energy of the localized bound state can be significantly different from the original value $\epsilon_l$. And even though $\epsilon_l-\hbar\Delta\omega$ is in the continuous energy band, the localized bound state still survives.

An interesting question is that if there is no localized bound state in $u_0(t,t_0)$, can strong driving generate one? Our numerical study shows no exception that strong driving cannot generate localized bound state(s) from none. 
This is actually a universal result. The argument is as follows. For the case that no localized bound state exists in $u_0(t,t_0)$, the time-dependent Hamiltonian in Eq.~\eqref{eq_Htot} can be reduced as
\begin{align}
H_{\rm tot}(t)\!=\!& \sum_k \epsilon_k c_k^\dag c_k \!+\! \Delta\epsilon_d(t) \sum_{k,k'} \lambda_{k k'} c^\dag_{k} c_{k'} \,. 
\end{align}
The energy shift of mode-$k$ is determined by its total coupling with the other modes, which is roughly characterized by $\sum_{k'} |\lambda_{kk'}|^2$.  Because there is no distinguished mode $c_k$, the energy shifts due to the existence of the driving are continuous with respect to $\epsilon_k$. This distortion of the energy band cannot transform the continuous energy band to be discontinuous. Therefore, no time-dependent localized bound state can be repelled out of the band.

It is noteworthy that the above case is different from the impact of $\overline{\epsilon}_d$ in
\begin{align}
H\!=& (\epsilon_s\!+\!\overline{\epsilon}_d) b^\dag b  \!+\!\sum_k \epsilon_k b_k^\dag b_k \!+\!\sum_k (V_{k} b^\dag b_k \!+\! h.c.)\,,
\end{align}
where there is a distinguished mode $b$. The energy shift of the system mode $b$ is determined by its total coupling with all the environmental modes, i.e., $\sum_{k}|V_k|^2$, which is a non-vanishing value. However, the energy shift of the environmental mode $b_k$ is determined by the value of $|V_k|^2$, which is vanishingly small for the continuous environmental structure. This results in the fact that due to \blk the interaction between the system and the environment, the energy shift of the system mode can be significant while the continuous energy band of the environment keeps identical to that of the free environment. % the energy shift of the system mode $b$ is determined by its total coupling with other modes, i.e., $\sum_{k}|V_k|^2$, which is a non-vanishing value.
As a consequence, it gives the possibility of generating one or several localized bound states in the total system even though the value of $\epsilon_s$ is in the continuous energy band. % In the static case such as the Hamiltonian considered in Eq.~\eqref{eq_H}, strong coupling between the system and the environment may lead to the generation of localized bound state even though the value of $\epsilon_s$ is in the continuous energy band. The driving field induces the energy shifts and transitions among the energy levels in the continuous band. 

%There are some time intervals that the open system possesses instantaneous localized bound state. However, the instants that no instantaneous localized bound state exists would inevitably make the particle in the system dissipate into the environment. Even though the open system possesses instantaneous localized bound state at an arbitrary instant, $u(t,t_0)$ would finally decay to zero. 

%When $\frac{2\pi}{T}<B$, where $B$ stands for the bandwidth, then the localized bound state always decays~\cite{CAL15}. 

\section{Conclusion}
\label{sec5}

In this work, we study the exact dynamics of open quantum systems to the case with periodic driving field. We present a general viewpoint of energy transfer in dealing with the controlling of decoherence dynamics and analyze the general properties of the decoherence dynamics in existence of the periodic driving. The driving field can be decomposed into the static part and the zero-averaged oscillating part, which play different roles in affecting the decoherence. %One is the static part, the other is the zero-averaged oscillating part. 
The static part of the driving can be constructive, destructive, or sustaining to the localized bound states avoiding decoherence, while the zero-averaged part can only be destructive or sustaining. % the static part of the driving can either generate or destroy localized bound states, while the zero-averaged part of the driving may preserve or destroy the existed localized bound states but cannot generate one from none. 
More precisely, if the original system possesses no localized bound state, the periodic driving field with zero average cannot generate the dissipationless localized bound state from nothing. If the initial system possesses localized bound state(s), 
the driving field may destroy it (them).

The destroying process of a localized bound state by the driving field is mediated by exchanging energy with the continuous energy band via the (zero-averaged periodic) driving. % With the driving field, particle in the localized bound state may absorb or release energy from the field and transit to the state in the continuous band. If this process can happen, then the localized bound state would be destroyed. If it cannot happen, then the system would have periodic localized bound state. 
In the weak driving case, the picture of energy transfer works nearly precisely. It is shown that if a localized bound state can absorb or release one or several energy quanta (which is the high-order process) from the driving field and transfer to the continuous energy band, i.e., $\epsilon_l\pm \hbar\omega_n \!\in \!\mathcal{B}$ or $\epsilon_l\pm n \hbar\Delta\omega \!\in \! \mathcal{B}$, where $\mathcal{B}$ stands for the continuous energy band, then it would inevitably dissipate. Otherwise, it would only be modulated as a time-dependent localized bound state.  \blk %High-order processes can also take part in the dissipation process. That is, if the localized bound state and the continuous energy band can exchange energy through several energy quanta of the fundamental frequency, the dissipation process also inevitably happens except that the decay rate is much smaller than the direct transition. 
For the strong driving case, the picture of energy transfer still works but should be modified. The energy shift of the localized bound state due to the energy renormalization is not small, making the condition of dissipation in the weak coupling case break down. 

The localized bound state of open quantum systems is decoherence-free in absence of the external driving field, offering its potential application of working as the quantum memory. Our study reveals that its decoherence is sensitively related to the frequency of the driving field, especially when the fundamental frequency $\Delta\omega$ satisfies that $\epsilon_l\pm \hbar\Delta\omega$ is around the edge of the environmental energy band. This offers us a method of determining whether to 
keep or erase the quantum information stored \blk in the localized bound state.

\acknowledgments
We thank Lian-Ao Wu, Yu-Wei Huang, and Yi Yang for the suggestions in dealing with the figures. We also thank Hon-Lam Lai for helpful discussions when identifying the problem. This work is supported by the Ministry of Science and Technology of the Republic of China under the Contracts No.~MOST-108-2112-M-006-009-MY3.

%%%%%%%%%%%%%%%%%%%%%%%%%%%%%%%%%%%%%%%%
% choose a .bib file
\bibliographystyle{apsrev4-1}
\bibliography{references}
%%%%%%%%%%%%%%%%%%%%%%%%%%%%%%%%%%%%%%%%

\end{document}